\documentclass{article}
\usepackage{amssymb}
\usepackage{amsmath}
\usepackage{harvard}

\input{tcilatex}

\begin{document}

\title{\textbf{Non-linear hydrodynamics of incommensurate intergrowth
compounds and quasicrystals}}
\author{\textbf{Paolo Maria Mariano}$%
{{}^\circ}%
^{\ast }$ \\
$%
{{}^\circ}%
$ Dipartimento di Ingegneria Strutturale e Geotecnica,\\
Universit\`{a} di Roma "La Sapienza",\\
via Eudossiana 18, I-00184 Roma (Italy),\\
e-mail: paolo.mariano@uniroma1.it,\\
$^{\ast }$now at Universit\`{a} del Molise, Campobasso (Italy).}
\maketitle

\begin{abstract}
Hamiltonian structures for non-linear hydrodynamics of incommensurate
intergrowth compounds (IIC) and quasicrystals (IQ) are constructed. We
discuss also the way to account for internal friction of phason nature. We
show that the existence of a self-force in IIC and IQ is not only matter of
constitutive issues, rather it is related with questions of $SO\left(
3\right) $ invariance. The covariant mechanics of discontinuity surfaces in
quasiperiodic structures is also analyzed. The attention is mainly focused
on the interaction between `diffuse' grain boundaries and sharp
discontinuity (moving possibly) surfaces.

\ \ \ \ \ \ \ 

\emph{To T. Y. for her moral vigour to oppose racialism of stupid people and
to face difficulties.}
\end{abstract}

\section{Introduction}

Quasiperiodic metallic alloys display two types of low energy excitations in
the hydrodynamic range: the standard \emph{phonon} modes associated with the
congruent distortions between neighboring material elements, and \emph{phason%
} modes (Goldstone degrees of freedom) due to local rearrangements of atomic
clusters. The former modes are represented by the standard displacement
field $\mathbf{\tilde{u}}$, while the latter ones by the phason vector field 
$\mathbf{\tilde{w}}$.

Here our attention is focused on incommensurate intergrowth compounds (IIC)
and proper icosahedral quasicrystals (IQ). For IIC, the phason displacement
field describes collective atomic modes associated with relative
displacements between incommensurate sublattices determining
quasiperiodicity. Basically, IIC can be considered as the result of
modifications of periodic structures. On the contrary, as shown by the 1984
experiments of D. Shechtman, I. Bleck, D. Gratias and J. W. Cahn [S], there
exist intrinsically quasiperiodic crystals (IQ) that are not modulation of
multiply twinned periodic crystalline structures or the composition of more
species of them. They admit icosahedral phases with consequent long-range
orientational order and absence of translational one. They violate the \emph{%
crystallographic restriction} prescribing that a crystal displays
periodicity and "cannot have certain forbidden symmetries, such as fivefold
rotation" [L] but do not constitute a new state of matter. For IQ, the
phason displacements describes substructural changes of diffusive nature
[RoLo]: (\emph{i}) collective atomic modes and (\emph{ii}) tunneling of
atoms below energetic barriers separating places at a distance lesser than
the atomic diameter.

Phason activity thus exists in IIC and IQ; however, the energetic landscape
is different. In the diffraction scenarios obtained by x-ray scattering
experiments, diffuse scattering is registered around Bragg peaks in both
cases. In principle, such a scattering could be represented through the
singular part of certain autocorrelation measures [BH]. However, for IIC
there are six sound-like branches while in the case of IQ just three
sound-like branches appear. As a consequence, kinetic energy can be
attributed to the phason activity in the case of IIC, while for IQ there may
be a sort of internal friction leading to viscous-like evolution because
phason activity displays diffusive nature. Since in the case of IIC, at each
point $\mathbf{X}$, the vector $\mathbf{w=\tilde{w}}\left( \mathbf{X}\right) 
$ represents the \emph{relative} displacement of incommensurate sublattices,
it is a measure of deformation and does enter the structure of the free
energy together with its gradient $\nabla \mathbf{w}$. It does not happen to
IQ where just $\nabla \mathbf{w}$ appears in the list of constitutive
entries of the energy as a representative of phason behavior. The explicit
dependence on $\nabla \mathbf{w}$\ varies according to the circumstance that
IQ is in a `locked phase' (i.e. a phase without phason contribution of
Debye-Waller type) or in an `unlocked phase' (see [JS]).

Taking into account analogies and differences, we present here a Hamiltonian
formalism for non-linear hydrodynamics of IIC and IQ. We follow strictly the
general framework of multifield theories [GC], [C], [M], [CM] unifying a
wide class of models of condensed matter with complex substructural
morphology. In discussing the non-linear elasticity of quasiperiodic
structures, we follow in particular the general results in [CM] dealing with
field theories in which the Hamiltonian accounts for order parameters taking
values on an abstract manifold. Here, at each point $\mathbf{X}$, our order
parameter is the phason displacement $\mathbf{w=\tilde{w}}\left( \mathbf{X}%
\right) $, belonging to some copy of the translation space $\mathbb{V}_{w}$
of the three-dimensional Euclidean point space $\mathcal{E}^{3}$.

The mechanics of quasiperiodic crystalline structures in the hydrodynamic
range, the one of elasticity and plasticity, has been discussed variously in
scientific literature. The attention has been focused primarily on
quasicrystals rather than on IIC starting from the work of T. C. Lubesky, S.
Ramaswamy, J. Toner [LRT] on (see [HWD], [RT], [DP], [JS], [RL], [RoLo],
[DYHW], [MSA]).

Here, we analyze the matter. Accomplishments are briefly summarized below.

\begin{description}
\item[(i)] Hamiltonian structures for both IIC and IQ are constructed. The
covariance of the balance of phason interactions follow. A balance of
interactions occurring when defects in the quasiperiodic structure are
permuted is then deduced.

\item[(ii)] It is possible to show that the existence of a self-force of
phason nature \emph{within }each material element is a consequence of $%
SO\left( 3\right) $ invariance. It is characterized by constitutive
instances: in the case of IQ it vanishes at thermodynamical equilibrium. A
non-standard integral balance of moments follows naturally from $SO\left(
3\right) $ invariance. It is different from the one postulated commonly (see
[HWD]) and gives rise (by localization) to the pointwise balance of phason
interactions that is not necessarily associated with an integral balance of
phason momentum.

\item[(iii)] We discuss also non conservative issues to account for the
internal friction in IQ. Our treatment contains the `minimal model'
presented by S. B. Rochal and V. L. Lorman in [RoLo].

\item[(iv)] Finally we describe the influence of phason activity on the
evolution of discontinuity surfaces in quasiperiodic crystalline structures.
We show covariance of the balances of phonon and phason interactions at the
discontinuity surfaces.
\end{description}

For the sake of simplicity our treatment deals only with isothermal
processes.

Algorithms for analyzing numerically specific cases follow naturally.

Unaspected phenomena may be evidenced. For example, in the linear case, due
to uncertainties in the experimental evaluation of phonon-phason coupling
coefficient in icosahedral quasicrystals, the combined use of Monte-Carlo
and finite element techniques allows us to put in evidence the possible
stochastic clustering of phonon and phason modes around macroscopic defects
like cracks. The relevant results will be presented in a forthcoming paper
with M. Gioffr\'{e} and F. L. Stazi.

\ \ \ \ \ \ \ \ \ \ \ \ \ \ \ \ \ 

\emph{Some notations}. For any pair of vector spaces $A$ and $B$ (with duals 
$A^{\ast }$ and $B^{\ast }$), $Hom\left( A,B\right) $ is the space of linear
maps from $A$ to $B$. For any manifold $M$, $T_{m}M$ is the tangent space of 
$M$ at $m\in M$, while $T_{m}^{\ast }M$ the relevant cotangent space.
Moreover, $Aut\left( A\right) $ indicates the space of automorphisms of $A$.
We will make use of two different regular bounded regions of the
three-dimensional Euclidean point space $\mathcal{E}^{3}$, namely $\mathcal{B%
}_{0}$\ and $\mathcal{B}$, and of two different copies\ $\mathbb{V}_{u}$ and 
$\mathbb{V}_{w}$ of the translation space of $\mathcal{E}^{3}$ (we may also
identify them as copies of $\mathbb{R}^{3}$). Capital letters $A,B,C...$
used as indices denote coordinates in $\mathcal{B}_{0}$, while $i,j,k...$
coordinates in $\mathcal{B}$.\ The differential operators $Div$ and $\nabla $
indicate respectively divergence and gradient calculated with respect to
coordinates in $\mathcal{B}_{0}$\ while $div$ and $grad$ are their
counterparts with respect to coordinates in $\mathcal{B}$. The superscript $%
T $ means transposition. The symbol $\partial _{y}$ means partial derivative
with respect to the entry "$y$". We indicate with the term \emph{part} any
subset of $\mathcal{B}_{0}$\ with non-vanishing volume and the same
regularity properties of $\mathcal{B}_{0}$. Let $\Sigma $\ be any smooth
surface in $\mathcal{B}_{0}$\ oriented by the normal $\mathbf{m}$\ at each
point, for any field $e\left( \cdot \right) $ defined on $\mathcal{B}_{0}$\
and differentiable there, we indicate with $\nabla _{\Sigma }$\ its surface
gradient along $\Sigma $, namely $\nabla _{\Sigma }e\left( \mathbf{X}\right)
=\nabla e\left( \mathbf{X}\right) \left( \mathbf{I-m\otimes m}\right) $,
with $\mathbf{I}$ the second-order unit tensor. The trace of $\nabla
_{\Sigma }e$\ is the surface divergence of $e$, namely $Div_{\Sigma }e$.\
Other notations will be explained later.

\section{Configurations, observers and relabeling}

\subsection{Configurations}

Our analysis deals with a quasiperiodic crystalline body occupying in its
reference place a regular\footnote{$\mathcal{B}_{0}$ is regular in the sense
of D-regions defined in [D].} region $\mathcal{B}_{0}$\ of the
three-dimensional Euclidean point space $\mathcal{E}^{3}$. A generic point $%
\mathbf{X}\in \mathcal{B}_{0}$ is identified with the centre of mass of a
crystalline cell (which is the characteristic material element) that one may
imagine collapsed at $\mathbf{X}$ in a coarse grained representation of the
quasiperiodic structure.

A standard deformation of the body is represented by a sufficiently smooth
injective mapping $\mathcal{B}_{0}\ni \mathbf{X}\overset{\mathbf{\tilde{x}}}{%
\longmapsto }\mathbf{x=\tilde{x}}\left( \mathbf{X}\right) \in \mathcal{E}%
^{3} $. The current place $\mathcal{B}=\mathbf{\tilde{x}}\left( \mathcal{B}%
_{0}\right) $\ of the body is a regular region too. The placement map $%
\mathbf{\tilde{x}}$ is also orientation preserving: at each $\mathbf{X}$ its
gradient $\mathbf{F}=\nabla \mathbf{x}$, i.e. the value of the field $%
\mathcal{B}_{0}\ni \mathbf{X}\overset{\mathbf{\tilde{F}}}{\longmapsto }%
\mathbf{F=\tilde{F}}\left( \mathbf{X}\right) \in Hom\left( T_{\mathbf{X}}%
\mathcal{B}_{0},T_{\mathbf{x}}\mathcal{B}\right) $, has positive determinant.

Let $\mathbf{g}$ be the spatial metric in $\mathcal{B}$ and $\gamma $\ the
metric in $\mathcal{B}_{0}$. The linear operator $\mathbf{F}^{T}\mathbf{F=C}%
\in Hom\left( T_{\mathbf{X}}\mathcal{B}_{0},T_{\mathbf{x}}^{\ast }\mathcal{B}%
\right) $ is the pull-back at $\mathbf{X}$ of $\mathbf{g}$ through $\mathbf{%
\tilde{x}}$, i.e., in coordinates, $C_{AB}=F_{A}^{Ti}g_{ij}F_{B}^{j}$. Then,
the difference $\left( \mathbf{C-\gamma }\right) $ is twice the non-linear
deformation tensor $\mathbf{E}$.

If we consider each material element as a perfect crystalline cell, during a
motion $\mathcal{B}_{0}\times \left[ 0,\bar{t}\right] \ni \left( \mathbf{X,}%
t\right) \overset{\mathbf{\tilde{x}}}{\longmapsto }\mathbf{x=\tilde{x}}%
\left( \mathbf{X,}t\right) \in \mathcal{E}^{3}$, the standard displacement
field $\mathbf{u=\tilde{u}}\left( \mathbf{X,}t\right) =\mathbf{\tilde{x}}%
\left( \mathbf{X,}t\right) -\mathbf{X}\in \mathbb{V}_{w}$ is the descriptor
of \emph{phonon degrees of freedom}. When the material element undergoes at
least one of the substructural changes $(i)$ and $(ii)$ described above,
namely collective atomic modes or tunneling of atoms, a sort of internal
shift occurs and is represented by a vector $\mathbf{w}$ so that we have a
vector field $\mathcal{B}_{0}\ni \mathbf{X}\overset{\mathbf{\tilde{w}}}{%
\longmapsto }\mathbf{w=\tilde{w}}\left( \mathbf{X}\right) \in \mathbb{V}_{w}$
that we presume sufficiently smooth over the body. During a motion, we then
have $\mathcal{B}_{0}\times \left[ 0,\bar{t}\right] \ni \left( \mathbf{X,}%
t\right) \overset{\mathbf{\tilde{w}}}{\longmapsto }\mathbf{w=\tilde{w}}%
\left( \mathbf{X,}t\right) \in \mathbb{V}_{w}$, with a slight abuse of
notation.

From the point of view of the general setting of \emph{multifield theories},
the copy $\mathbb{V}_{w}$ of the translation space $\mathbb{V}$\ over $%
\mathcal{E}^{3}$, containing $\mathbf{w}$, plays the r\^{o}le of the
manifold of morphological descriptors (order parameters) of the material
substructure [C], [M].

One may consider (see [HWD]) a global displacement $\mathbf{\bar{u}}$
belonging to $\mathbb{V}_{u}\mathbb{\oplus V}_{w}$. By indicating with $%
\mathbf{x}^{\prime }=\mathbf{\tilde{x}}^{\prime }\left( \mathbf{X}\right) $
the point given by $\mathbf{x}^{\prime }=\mathbf{X+\bar{u}=X+u+w}$, and with 
$\mathbf{F}^{\prime }$ the gradient $\nabla \mathbf{\tilde{x}}^{\prime
}\left( \mathbf{X}\right) $, we get additive and multiplicative
decompositions given respectively by $\mathbf{F}^{\prime }=\mathbf{F}+\nabla 
\mathbf{w}$ and $\mathbf{F}^{\prime }=\mathbf{F}^{ph}\mathbf{F}$, with $%
\mathbf{F}^{ph}=\mathbf{I}+\left( \nabla \mathbf{w}\right) \mathbf{F}^{-1}$.
Really, since $\mathbf{\tilde{x}}$ is one-to-one, one may construct a
representation of $\mathbf{w}$ on the `apparent' current place $\mathcal{B}$
of the body. By indicating with $\mathbf{w}_{a}=\mathbf{\tilde{w}}_{a}\left( 
\mathbf{x}\right) $ the image of $\mathbf{w}$ attached at $\mathbf{x=\tilde{x%
}}\left( \mathbf{X}\right) \in \mathcal{B}$, we get $\mathbf{\tilde{w}}_{a}=%
\mathbf{\tilde{w}}$ $\circ $ $\mathbf{\tilde{x}}^{-1}$ so that $\left(
\nabla \mathbf{w}\right) \mathbf{F}^{-1}=grad\mathbf{w}_{a}$. As a
consequence, the interpretation of the multiplicative decomposition of $%
\mathbf{F}$ may be the following: we may deform first the body at a coarse
grained level maintaining frozen phason activity, then we may allow
collective atomic modes to develop. In other words, by indicating with $%
\mathbf{\tilde{x}}^{ph}$ the mapping $\mathbf{\tilde{x}}^{ph}=\mathbf{\tilde{%
x}}^{\prime }\circ \mathbf{\tilde{x}}^{-1}$, we see that $\mathbf{F}^{ph}=%
\mathbf{I}+grad\mathbf{w}_{a}$ is the gradient of deformation from $\mathcal{%
B}$ to $\mathbf{\tilde{x}}^{ph}\left( \mathcal{B}\right) $, namely there is
a piecewise continuous map $\mathbf{\tilde{F}}^{ph}$ such that $\mathcal{B}%
_{0}\ni \mathbf{X}\overset{\mathbf{\tilde{F}}^{ph}}{\longmapsto }\mathbf{F}%
^{ph}=\mathbf{\tilde{F}}^{ph}\left( \mathbf{X}\right) \in Hom\left( T_{%
\mathbf{x}}\mathcal{B},T_{\mathbf{x}^{\prime }}\mathbf{\tilde{x}}^{ph}\left( 
\mathcal{B}\right) \right) $. The map $\mathbf{\tilde{x}}^{\prime }$\
describes the circumstance that collective atomic modes or tunneling of
atoms occurring within each crystalline cell may shift the centre of mass of
the crystalline cell itself from its current (in certain sense `apparent')
place $\mathbf{x}$.

Finally, we indicate with $\mathbf{\dot{x}}=\frac{d}{dt}\mathbf{\tilde{x}}%
\left( \mathbf{X,}t\right) $ and $\mathbf{\dot{w}}=\frac{d}{dt}\mathbf{%
\tilde{w}}\left( \mathbf{X,}t\right) $ rates in the reference description
and might use also $\mathbf{\dot{u}}=\frac{d}{dt}\mathbf{\tilde{u}}\left( 
\mathbf{X,}t\right) $ instead of $\mathbf{\dot{x}}$ to put in evidence the r%
\^{o}le of phonon and phason degrees of freedom.

If we restrict our attention to infinitesimal deformation regime in which $%
\mathcal{B}$\ can be `confused' with $\mathcal{B}_{0}$ in the sense that $%
\mathbf{\dot{x}\approx u}$ at each $\mathbf{X}$, in addition to the standard
compatibility condition $curl$ $curl$ $sym\nabla \mathbf{u=0}$, we get also a%
\emph{\ phason compatibility condition }$curl$ $curl$ $\nabla \mathbf{w=0}$
that would imply eventually an energetic contribution of phason spin.

\subsection{Observers and relabeling}

For the mechanics of quasicrystals the definition of the concept of observer
follows general issues of the mechanics of complex materials (see [M04])
involving in such a definition the representation of all geometrical
environments necessary to the description of the material morphology.

Three sets enter in fact the geometrical picture of a quasi-periodic
crystalline body: the point space $\mathcal{E}^{3}$ (i.e. the standard
ambient space), the translation space $\mathbb{V}_{w}$ (containing phason
degrees of freedom) and the interval of time $\left[ 0,\bar{t}\right] $. An
observer $\mathcal{O}$\ is then a representation of $\mathcal{E}^{3}$, $%
\mathbb{V}_{w}$ and $\left[ 0,\bar{t}\right] $.

We consider also relabeling of material elements in $\mathcal{B}_{0}$,
simulating a redistribution of possible defects.

\textbf{Relabeling}. Formally, a `permutation of inhomogeneities' in $%
\mathcal{B}_{0}$ is described by the action of the special group of isocoric
diffeomorphisms $SDiff$ on $\mathcal{B}_{0}$. So that we have a map

\begin{itemize}
\item $\mathbb{R}^{+}\ni s_{1}\longmapsto \mathbf{f}_{s_{1}}^{1}\in
SDiff\left( \mathcal{B}_{0}\right) $, with $\mathbf{f}_{0}^{1}$ the identity.
\end{itemize}

At each $s_{1}$ we get $\mathbf{X\longmapsto f}_{s_{1}}^{1}\left( \mathbf{X}%
\right) $, with $Div\mathbf{f}_{s_{1}}^{1\prime }\left( \mathbf{X}\right) =0$%
, where the prime denotes differentiation with respect to the parameter $%
s_{1}$. We put $\mathbf{f}_{0}^{1\prime }\left( \mathbf{X}\right) =\mathfrak{%
w}$.

\textbf{Changes of observers}. We consider observers agreeing about the
measure of time so that a generic change of observer involves just a couple
of transformations: one of the ambient space $\mathcal{E}^{3}$, the other of 
$\mathbb{V}_{w}$. They are described by the parametrized families of
mappings defined below.

\begin{itemize}
\item $\mathbb{R}^{+}\ni s_{2}\longmapsto \mathbf{f}_{s_{2}}^{2}\in
Aut\left( \mathcal{E}^{3}\right) $, with $\mathbf{f}_{0}^{2}$ the identity.
We put $\mathbf{f}_{0}^{2\prime }\left( \mathbf{X}\right) =\mathbf{v}$.

\item A Lie group $G$, with Lie algebra $\mathfrak{g}$, acts over $\mathbb{V}%
_{w}$. If $\xi \in \mathfrak{g}$, its action over $\mathbf{w\in }\mathbb{V}%
_{w}$ is indicated with $\xi _{\mathbb{V}_{w}}\left( \mathbf{w}\right) $. By
indicating with $\mathbf{w}_{g}$ the value of $\mathbf{w}$ after the action%
\footnote{%
It is not essential to render precise if the action is from the left or from
the right.} of $g\in G$, if we consider a one-parameter smooth curve $%
\mathbb{R}^{+}\ni s_{3}\longmapsto g_{s_{3}}\in G$ over $G$ such that $\xi =%
\frac{dg_{s_{3}}}{ds_{3}}\left\vert _{s_{3}=0}\right. $\ and its
corresponding orbit $s_{1}\longmapsto \mathbf{w}_{g_{s_{3}}}$\ over $\mathbb{%
V}_{w}$, starting from a given $\mathbf{w}$, we have $\xi _{\mathbb{V}%
_{w}}\left( \mathbf{w}\right) =\frac{d}{ds_{3}}\mathbf{w}_{g_{s_{3}}}\left%
\vert _{s_{3}=0}\right. $.
\end{itemize}

\section{Lagrangian structures for phonon-phason elasticity}

Up to this point just geometry has been involved. In constructing a
mechanical model of a body, after the description of its morphology, one
discusses the representation of interactions and their balance first, then
the explicit representation of constitutive relations. The two issues are
essentially separated. The representation of interactions by means of
appropriate vectors or higher order tensors is a consequence of the
essential geometrical description of the body (interactions are in fact
entities power conjugated with the rates of morphological descriptors) and
the balance is independent of the constitutive nature of the material.

When we develop Lagrangian and Hamiltonian formalisms as below, in
introducing the Lagrangian density just after geometrical issues, we put on
the same ground the representation of interactions and constitutive issues
because they are mixed in the variational description.

Let us consider a fiber bundle%
\begin{equation}
\pi :\mathcal{Y\rightarrow }\mathcal{B}_{0}\times \left[ 0,\bar{t}\right]
\label{1}
\end{equation}%
such that $\pi ^{-1}\left( \mathbf{X,}t\right) =\mathcal{E}^{3}\times 
\mathbb{V}_{w}$ is the prototype fiber. A generic section $\eta \in \Gamma
\left( \mathcal{Y}\right) $ is then a mapping $\eta :\mathcal{B}_{0}\times %
\left[ 0,\bar{t}\right] \longrightarrow \mathcal{Y}$ such that $\eta \left( 
\mathbf{X,}t\right) =\left( \mathbf{X},t,\mathbf{x,w}\right) $ with $\mathbf{%
x}$ and $\mathbf{w}$ in the fiber $\pi ^{-1}\left( \mathbf{X,}t\right) $. If
sufficient smoothness for sections is allowed, the first jet bundle $J^{1}%
\mathcal{Y}$ over $\mathcal{Y}$ is such that%
\begin{equation}
J^{1}\mathcal{Y}\ni j^{1}\left( \eta \right) \left( \mathbf{X,}t\right)
=\left( \mathbf{X},t,\mathbf{x,\dot{x},F,w,\dot{w},}\nabla \mathbf{w}\right)
.
\end{equation}

In the conservative case we presume that the canonical Lagrangian $3+1$ form%
\begin{equation}
L:J^{1}\mathcal{Y\rightarrow \wedge }^{3+1}\left( \mathcal{B}_{0}\times %
\left[ 0,\bar{t}\right] \right)
\end{equation}%
admits a sufficiently smooth density $\mathcal{L}$\ such that%
\begin{equation}
L\left( j^{1}\left( \eta \right) \left( \mathbf{X,}t\right) \right) =%
\mathcal{L}\left( \mathbf{X},t,\mathbf{x,\dot{x},F,w,\dot{w},}\nabla \mathbf{%
w}\right) d\left( vol\right) \mathbf{\wedge }dt.
\end{equation}%
with $\mathcal{L}$\ defined by%
\begin{equation*}
\mathcal{L}\left( \mathbf{X},t,\mathbf{x,\dot{x},F,w,\dot{w},}\nabla \mathbf{%
w}\right) =\frac{1}{2}\rho _{0}\left\vert \mathbf{\dot{x}}\right\vert ^{2}+%
\frac{1}{2}\bar{\rho}\left\vert \mathbf{\dot{w}}\right\vert ^{2}-
\end{equation*}%
\begin{equation}
-\rho _{0}e\left( \mathbf{X,F,w,}\nabla \mathbf{w}\right) -\rho _{0}w\left( 
\mathbf{x}\right) ,
\end{equation}%
where $\rho _{0}$\ is the referential mass density (conserved during the
motion), $\bar{\rho}$\ an inertia coefficient for possible phason kinetics
(see [HDW], [RL]), $e$ the elastic energy density and $w$\ the density of
the potential of external actions, all per unit mass. Here we do not
consider possible bulk external direct actions on phason changes.

We then evaluate the variation of the total Lagrangian $\bar{L}\left( 
\mathcal{B}_{0}\right) $ given by $\bar{L}\left( \mathcal{B}_{0}\right)
=\int_{\mathcal{B}_{0}\times \left[ 0,\bar{t}\right] }\mathcal{L}d\left(
vol\right) \wedge dt$ and we may find at least one section (with the
properties of $\mathbf{\tilde{x}}$ and $\mathbf{\tilde{w}}$) satisfying
Euler-Lagrange equations for $\bar{L}\left( \mathcal{B}_{0}\right) $, namely%
\begin{equation}
\overset{\cdot }{\overline{\partial _{\mathbf{\dot{x}}}\mathcal{L}}}%
=\partial _{\mathbf{x}}\mathcal{L}-Div\partial _{\mathbf{F}}\mathcal{L},
\label{A}
\end{equation}%
\begin{equation}
\overset{\cdot }{\overline{\partial _{\mathbf{\dot{w}}}\mathcal{L}}}%
=\partial _{\mathbf{w}}\mathcal{L}-Div\partial _{\nabla \mathbf{w}}\mathcal{L%
}.  \label{B}
\end{equation}

\textbf{Definition 1} (invariance of\textbf{\ }$\mathcal{L}$). $\mathcal{L}$
is \emph{invariant} with respect to the action of $\mathbf{f}_{s_{1}}^{1}$, $%
\mathbf{f}_{s_{2}}^{2}$ and $G$ if%
\begin{equation*}
\mathcal{L}\left( \mathbf{X,x,\dot{x},F,w,\dot{w},}\nabla \mathbf{w}\right) =
\end{equation*}%
\begin{equation}
=\mathcal{L}\left( \mathbf{f}^{1}\mathbf{,f}^{2}\mathbf{,}\left( grad\mathbf{%
\mathbf{f}}^{2}\right) \mathbf{\dot{x},}\left( grad\mathbf{\mathbf{f}}%
^{2}\right) \mathbf{F}\left( \nabla \mathbf{f}^{1}\right) ^{-1}\mathbf{,w}%
_{g},\mathbf{\dot{w}}_{g},\left( \nabla \mathbf{w}_{g}\right) \left( \nabla 
\mathbf{f}^{1}\right) ^{-1}\right) .
\end{equation}%
where we indicate with $\mathbf{f}^{1}$, $\mathbf{f}^{2}$ and $\mathbf{w}%
_{g} $\ the values $\mathbf{f}_{s_{1}}^{1}\left( \mathbf{X}\right) $, $%
\mathbf{f}_{s_{2}}^{2}\left( \mathbf{x}\right) $, $\mathbf{w}%
_{g_{s_{3}}}\left( \mathbf{X}\right) $.

\ \ \ \ \ \ \ 

Let $\mathcal{Q}$\ and $\mathfrak{F}$\ be scalar and vector densities given
respectively by%
\begin{equation}
\mathcal{Q}=\partial _{\mathbf{\dot{x}}}\mathcal{L\cdot }\left( \mathbf{v}-%
\mathbf{F}\mathfrak{w}\right) +\partial _{\mathbf{\dot{w}}}\mathcal{L\cdot }%
\left( \xi _{\mathbb{V}_{w}}\left( \mathbf{w}\right) \mathbf{-}\left( \nabla 
\mathbf{w}\right) \mathfrak{w}\right) ,
\end{equation}%
\begin{equation}
\mathfrak{F}=\mathcal{L}\mathfrak{w}\mathbf{+}\left( \partial _{\mathbf{F}}%
\mathcal{L}\right) ^{T}\left( \mathbf{v}-\mathbf{F}\mathfrak{w}\right)
+\left( \partial _{\nabla \mathbf{w}}\mathcal{L}\right) ^{T}\left( \xi _{%
\mathbb{V}_{w}}\left( \mathbf{w}\right) \mathbf{-}\left( \nabla \mathbf{w}%
\right) \mathfrak{w}\right) .
\end{equation}

\textbf{Theorem 1}. \emph{If the Lagrangian density} $\mathcal{L}$ \emph{is
invariant under} $\mathbf{f}_{s_{1}}^{1}$, $\mathbf{f}_{s_{2}}^{2}$\emph{\
and }$\emph{G}$\emph{, then}%
\begin{equation}
\mathcal{\dot{Q}}+Div\mathfrak{F}=0\text{.}  \label{Noe}
\end{equation}

Theorem above is a version for quasiperiodic bodies of Noether theorem. A
generalization of it for multifield theories that involve order parameters
belonging to abstract manifolds is proven in [CM].

\ \ \ \ \ \ \ \ \ \ \ \ 

\textbf{Corollary 1}. If $\mathbf{\mathbf{f}}_{s_{2}}^{2}$ alone acts on $%
\mathcal{L}$ leaving $\mathbf{v}$ \emph{arbitrary}, from (\ref{Noe}) we get
in covariant way the balance of phonon interaction (standard Cauchy's
balance of momentum)%
\begin{equation}
\rho _{0}\mathbf{\ddot{x}=}\rho _{0}\mathbf{b+}Div\mathbf{P},  \label{Cauchy}
\end{equation}%
where $\mathbf{P=-}\partial _{\mathbf{F}}\mathcal{L}$ is the first
Piola-Kirchhoff stress and $\mathbf{b}=\partial _{\mathbf{x}}\mathcal{L}$
the vector of body forces.

\ \ \ \ \ \ \ \ \ \ \ \ \ 

At each $\mathbf{X}$ in $\mathcal{B}_{0}$, $\mathbf{P}$ maps linearly
normals to surfaces through $\mathbf{X}$ into tensions at $\mathbf{x}$ in $%
\mathcal{B}$. There is then a map $\mathbf{\tilde{P}}$\ such that $\mathcal{B%
}_{0}\in \mathbf{X}\longmapsto \mathbf{P=\tilde{P}}\left( \mathbf{X}\right)
\in Hom\left( T_{\mathbf{X}}^{\ast }\mathcal{B}_{0},T_{\mathbf{x}}^{\ast }%
\mathcal{B}\right) $.

\ \ \ \ \ \ \ \ \ 

\textbf{Corollary 2}. If $G$ \emph{arbitrary} acts alone on $\mathcal{L}$,
from (\ref{Noe}) we obtain the balance of phason interactions%
\begin{equation}
\bar{\rho}\mathbf{\ddot{w}}=-\mathbf{z}+Div\mathcal{S},  \label{BSI}
\end{equation}%
in covariant way, where $\mathcal{S}=-\partial _{\nabla \mathbf{w}}\mathcal{L%
}$ represents \emph{phason stress} due to the relative influence of the
phason activity between neighboring material elements; $\mathbf{z=-}\rho
_{0}\partial _{\mathbf{w}}e$ (\emph{self-force}) describes self-interactions
of phason nature within each material element.

\ \ \ \ \ \ \ \ \ \ 

At each $\mathbf{X}$, $\mathcal{S}\in Hom\left( T_{\mathbf{X}}^{\ast }%
\mathcal{B}_{0},T_{\mathbf{w}}^{\ast }\mathbb{V}_{w}\right) $ and $\mathbf{z}%
\in T_{\mathbf{w}}^{\ast }\mathbb{V}_{w}$. The phason stress $\mathcal{S}$
maps linearly normals to surfaces through $\mathbf{X}$ in $\mathcal{B}_{0}$
into tensions of phason nature, i.e. elements of $T_{\mathbf{w}}^{\ast }%
\mathbb{V}_{w}\simeq \mathbb{R}^{3}$.

\ \ \ \ \ \ \ \ \ \ \ \ \ \ 

\textbf{Corollary 3}. Let $G=SO\left( 3\right) $ and, for any element $%
\mathbf{\dot{q}}\wedge $\ of its Lie algebra, $\mathbf{\mathbf{f}}%
_{s_{3}}^{2}$ be such that $\mathbf{v=\dot{q}}\wedge \left( \mathbf{x}-%
\mathbf{x}_{0}\right) $ with $\mathbf{x}_{0}$\ a fixed point in space. If $%
\mathcal{L}$ is independent of $\mathbf{x}$ and only the special choices of $%
\mathbf{\mathbf{f}}_{s_{2}}^{2}$ and $G$ just defined act on $\mathcal{L}$,
one gets from (\ref{Noe}) 
\begin{equation}
skw\left( \partial _{\mathbf{F}}e\mathbf{F}^{T}+\mathbf{w}\otimes \partial _{%
\mathbf{w}}e+\left( \nabla \mathbf{w}\right) \partial _{\nabla \mathbf{w}%
}e^{T}\right) =\text{0}\mathbf{,}  \label{Tor}
\end{equation}%
where $skw\left( \cdot \right) $ extracts the skew-symmetric part of its
argument.

\ \ \ \ \ \ \ \ 

\textbf{Corollary 4}. If $\mathbf{f}_{s_{1}}^{1}$ alone acts on $\mathcal{L}$%
, with $\mathfrak{w}$ \emph{arbitrary}, from (\ref{Noe}) one gets 
\begin{equation}
\overset{\cdot }{\overline{\left( \mathbf{F}^{T}\partial _{\mathbf{\dot{x}}}%
\mathcal{L}+\nabla \mathbf{\nu }^{T}\partial _{\mathbf{\dot{\nu}}}\mathcal{L}%
\right) }}-Div\left( \mathbb{P-}\frac{1}{2}\left( \rho _{0}\left\vert 
\mathbf{\dot{x}}\right\vert ^{2}+\bar{\rho}\left\vert \mathbf{\dot{w}}%
\right\vert ^{2}\right) \mathbf{I}\right) -\partial _{\mathbf{X}}\mathcal{L}=%
\mathbf{0}
\end{equation}%
where $\mathbb{P}=\rho _{0}e\mathbf{I}-\mathbf{F}^{T}\mathbf{P}-\nabla 
\mathbf{w}^{T}\mathcal{S}$, with $\mathbf{I}$ the second order unit tensor,
is a generalized Eshelby tensor accounting for phason activity (a special
case of the general one obtained in [M]).

\ \ \ \ \ \ \ \ \ \ \ 

\textbf{Corollary 5}. Let $G=SO\left( 3\right) $ and, for any element $%
\mathfrak{\dot{q}}\wedge $\ of its Lie algebra, $\mathbf{\mathbf{f}}%
_{s_{1}}^{1}$ is such that $\mathbf{w=}\mathfrak{\dot{q}}\wedge \left( 
\mathbf{X}-\mathbf{X}_{0}\right) $ with $\mathbf{X}_{0}$ a fixed point in $%
\mathcal{B}_{0}$. If the body is homogeneous, and only the special choices
of $\mathbf{\mathbf{f}}_{s_{2}}^{2}$ and $G$ just defined act on $\mathcal{L}
$, $\mathbb{P}$ is symmetric.

\ \ \ \ \ \ \ \ \ \ \ \ \ \ \ \ \ \ 

\textbf{Remark 1} (reduction to IQ). In the case of IQ, the elastic
potential $e$ does not depend on $\mathbf{w}$ and inertial effects
associated with phason modes are absent. So, in the purely conservative case
the balance of phason interactions and (\ref{Tor})\ become%
\begin{equation}
Div\mathcal{S}=0,\text{ \ \ }skw\left( \partial _{\mathbf{F}}e\mathbf{F}%
^{T}+\left( \nabla \mathbf{w}\right) \partial _{\nabla \mathbf{w}%
}e^{T}\right) =\text{0},
\end{equation}%
respectively.

\ \ \ \ \ \ \ \ \ \ \ \ \ \ \ \ 

\textbf{Remark 2}. Notice that for fixed $\mathbf{w}$ and $\nabla \mathbf{w}$
a standard result of non-linear elasticity must hold: namely $e\left( \cdot ,%
\mathbf{w,}\nabla \mathbf{w}\right) $ cannot be convex in $\mathbf{F}$ for
reasons of $SO\left( 3\right) $ invariance when large deformations occur.
The same property needs to be satisfied by $e\left( \cdot ,\nabla \mathbf{w}%
\right) $\ in the case of IQ.

\subsection{Universal phonon-phason changes in quasicrystals}

For an icosahedral quasicrystal (IQ) we say that the deformation is \emph{%
affine} when $\mathbf{F}^{\prime }$ does not depend on $\mathbf{X}$.

Moreover, if we consider deformations that can be controllable just by
applied macroscopic tractions, excluding in this way body forces, we call 
\emph{universal} all deformations that can occur in these conditions for all
bodies in a given class.

\ \ \ \ \ \ \ \ \ \ \ \ \ \ \ \ \ \ \ \ \ \ \ \ \ 

\textbf{Theorem 2}. \emph{Let the mappings} $\left( \mathbf{F,}\nabla 
\mathbf{w}\right) \overset{\mathbf{\tilde{P}}}{\longmapsto }\mathbf{P=\tilde{%
P}}\left( \mathbf{F,}\nabla \mathbf{w}\right) $ \emph{and} $\left( \mathbf{F,%
}\nabla \mathbf{w}\right) \overset{\mathcal{\tilde{S}}}{\longmapsto }%
\mathcal{S}=\mathcal{\tilde{S}}\left( \mathbf{F,}\nabla \mathbf{w}\right) $ 
\emph{admit bounded partial derivatives with respect to their entries and,
at each }$\mathbf{X}$\emph{, }%
\begin{equation}
\det \left( 
\begin{array}{cc}
\partial _{\mathbf{F}}\mathbf{P} & \partial _{\nabla \mathbf{w}}\mathbf{P}
\\ 
\partial _{\mathbf{w}}\mathcal{S} & \partial _{\nabla \mathbf{w}}\mathcal{S}%
\end{array}%
\right) \neq 0.  \label{det}
\end{equation}%
\emph{All universal static deformations of homogeneous (purely) elastic
quasicrystals satisfying the restriction }(\ref{det}) \emph{are affine.}

\ \ \ \ \ \ \ \ \ \ \ \ \ \ \ \ 

Such a theorem is in a certain sense a middle generalization of a standard
result in non-linear elasticity theory of simple bodies (see [A], p. 506).
Here the difference relies upon the circumstance that phason degrees of
freedom are involved.

To prove it, first recall that in absence of body forces and in conditions
of homogeneity, for IQ the equilibrium equations read%
\begin{equation}
Div\mathbf{P}=0,\text{ \ \ \ \ \ }Div\mathcal{S}=0.
\end{equation}%
We have also%
\begin{equation}
Div\mathbf{P}=\left( \partial _{\mathbf{F}}\mathbf{P}\right) \nabla \mathbf{F%
}+\left( \partial _{\nabla \mathbf{w}}\mathbf{P}\right) \nabla \nabla 
\mathbf{w}=0,
\end{equation}%
\begin{equation}
Div\mathcal{S}=\left( \partial _{\mathbf{w}}\mathcal{S}\right) \nabla 
\mathbf{F}+\left( \partial _{\nabla \mathbf{w}}\mathcal{S}\right) \nabla
\nabla \mathbf{w}=0,
\end{equation}%
i.e.%
\begin{equation}
\mathbb{A}_{1}\nabla \mathbf{F}+\mathbb{A}_{2}\nabla \nabla \mathbf{w}=0,
\label{21}
\end{equation}%
\begin{equation}
\mathbb{A}_{3}\nabla \mathbf{F}+\mathbb{A}_{4}\nabla \nabla \mathbf{w}=0,
\label{22}
\end{equation}%
with $\mathbb{A}_{i}$'s fourth-order tensors that are arbitrary because the
explicit form of the mappings $\mathbf{\tilde{P}}$ and $\mathcal{\tilde{S}}$
is not specified. They are also constant because the material is
homogeneous. Consequently, thanks to (\ref{det}) the solution to (\ref{21})
and (\ref{22}) provides $\nabla \mathbf{F}=0$ and $\nabla \nabla \mathbf{w}%
=0 $, that is $\mathbf{F}$ and $\nabla \mathbf{w}$ must be constant. As a
consequence, thanks to the additive decomposition $\mathbf{F}^{\prime }=%
\mathbf{F}+\nabla \mathbf{w}$, $\mathbf{F}^{\prime }$ is constant as well.\ 

\subsection{Mutations of material metric}

The extended Eshelby tensor $\mathbb{P}=\rho _{0}e\mathbf{I}-\mathbf{F}^{T}%
\mathbf{P}-\nabla \mathbf{w}^{T}\mathcal{S}$, accounting for phason effects,
enters the picture of the interactions involved in mutations (such as
evolution of defects, interfaces etc.) [M]. These mutations may be
represented by means of `alterations' of the geometrical structure of $%
\mathcal{B}_{0}$\ which, on the contrary, would remain fixed once and for
all. In particular, we focus here our attention just on mutations that may
involve changes in the material metric $\mathbf{\gamma }$ defined on $%
\mathcal{B}_{0}$\ and energy associated with them. Such a kind of situation
may occur in plastic flows (see [CM98]) or in the alteration of possible
pre-stressed states (see, e.g., [M1]). From now on we assume in this section 
$\rho _{0}=1$ for notational convenience.

We then consider (with some slight abuse of notation)\ a density of elastic
energy of the form%
\begin{equation}
e=\tilde{e}\left( \mathbf{\gamma },\mathbf{F,w,}\nabla \mathbf{w}\right)
\end{equation}%
in which we express explicitly the presence of $\mathbf{\gamma }$\ and
require that $\tilde{e}$\emph{\ is invariant under the action of the group
of point-valued diffeomorphisms defined on }$\mathcal{B}_{0}$ \emph{and
altering it}.

In other words, we require that $\tilde{e}$ is invariant under \emph{virtual}
superposition of sufficiently smooth deformations altering the reference
configuration. Notice that this kind of request of invariance is more than
the request of invariance with respect to relabeling because here it is not
required that diffeomorphisms involved are isocoric.

To this end, we then consider a one-parameter family $\mathbf{h}_{s}$ of
sufficiently smooth point-valued diffeomorphisms defined over $\mathcal{B}%
_{0}$\ and indicate with $\mathfrak{u}$ and $\mathbf{H}_{s}$ the derivatives 
$\frac{d}{ds}\mathbf{h}_{s}\left\vert _{s=0}\right. $ and $\nabla \mathbf{h}%
_{s}$, respectively. After the action of $\mathbf{h}_{s}$, the density $e$
changes as%
\begin{equation}
e\overset{\mathbf{h}_{s}}{\longmapsto }e_{\mathbf{h}_{s}}=\left( \det 
\mathbf{H}_{s}\right) \tilde{e}\left( \mathbf{H}_{s}^{-T}\mathbf{\gamma H}%
_{s}^{-1},\mathbf{FH}_{s}^{-1},\mathbf{w},\left( \nabla \mathbf{w}\right) 
\mathbf{H}_{s}^{-1}\right) .
\end{equation}

\ \ \ \ \ \ \ \ \ 

\textbf{Theorem 3}. \emph{If for }IIC \emph{and }IQ \emph{the energy density 
}$e$\emph{\ depends on the metric }$\mathbf{\gamma }$\emph{\ in }$\mathcal{B}%
_{0}$ \emph{and is invariant in the sense defined above, one gets}%
\begin{equation}
\mathbb{P}_{B}^{A}=2\left( \partial _{\mathbf{\gamma }}e\right) ^{AC}\gamma
_{CB}.  \label{RB}
\end{equation}

\ \ \ \ \ \ \ \ \ \ \ \ 

The proof relies upon the circumstance that the requirement of invariance of 
$e$ under the action of $\mathbf{h}_{s}$\ implies $\frac{d}{ds}e_{\mathbf{h}%
_{s}}\left\vert _{s=0}\right. $. It coincides with%
\begin{equation*}
\tilde{e}\left( \mathbf{\gamma },\mathbf{F,w,}\nabla \mathbf{w}\right)
tr\nabla \mathfrak{u}-\partial _{\mathbf{\gamma }}e\cdot \left( \left(
\nabla \mathfrak{u}\right) ^{T}\mathbf{\gamma }+\mathbf{\gamma }\left(
\nabla \mathfrak{u}\right) \right) -
\end{equation*}%
\begin{equation}
-\partial _{\mathbf{F}}e\cdot \mathbf{F}\nabla \mathfrak{u}-\partial
_{\nabla \mathbf{w}}e\cdot \left( \nabla \mathbf{w}\right) \nabla \mathfrak{u%
}=0
\end{equation}%
since $\frac{d}{ds}\mathbf{H}_{s}^{-1}\left\vert _{s=0}\right. =-\nabla 
\mathfrak{u}$. As a consequence of the symmetry of $\mathbf{\gamma }$, we
then get%
\begin{equation}
\left( e\mathbf{I}-\mathbf{F}^{T}\mathbf{P}-\nabla \mathbf{w}^{T}\mathcal{S}%
-2\left( \partial _{\mathbf{\gamma }}e\right) \mathbf{\gamma }\right) \cdot
\nabla \mathfrak{u}=0,
\end{equation}%
which implies (\ref{RB}), thanks to the arbitrariness of $\nabla \mathfrak{u}
$.

\textbf{Remark 3}. Notice that%
\begin{equation}
\mathbb{P}^{AD}=\mathbb{P}_{B}^{A}\gamma ^{BD}=2\left( \partial _{\mathbf{%
\gamma }}e\right) ^{AC}\gamma _{CB}\gamma ^{BD}=2\left( \partial _{\mathbf{%
\gamma }}e\right) ^{AD}
\end{equation}%
is symmetric since $\mathbf{\gamma }$ does.

\section{Elementary Hamiltonian structures}

Hamiltonian structures follow in a natural way from the Lagrangian
representation described so far. Let in fact $\mathbf{p}$ and $\mathbf{\mu }$
be respectively the \emph{canonical momentum} and the\ \emph{canonical
phason momentum} defined by $\mathbf{p=}\partial _{\mathbf{\dot{x}}}\mathcal{%
L}$ and $\mathbf{\mu =}\partial _{\mathbf{\dot{w}}}\mathcal{L}$.

The \emph{Hamiltonian density} $\mathcal{H}$ is then given by%
\begin{equation}
\mathcal{H}\left( \mathbf{X,x,p,F,w,\mu ,}\nabla \mathbf{\nu }\right) =%
\mathbf{p}\mathfrak{\cdot }\mathbf{\dot{x}}+\mathbf{\mu }\cdot \mathbf{\dot{w%
}}-\mathcal{L}\left( \mathbf{X,x,\dot{x},F,w,\dot{w},}\nabla \mathbf{w}%
\right) .
\end{equation}

In terms of partial derivatives of $\mathcal{H}$, the balances (\ref{A}) and
(\ref{B}) can be written as 
\begin{eqnarray}
\mathbf{\dot{p}} &\mathbf{=}&\mathbf{-}\partial _{\mathbf{x}}\mathcal{H}%
+Div\partial _{\mathbf{F}}\mathcal{H},  \notag \\
\mathbf{\dot{x}} &\mathbf{=}&\partial _{\mathbf{p}}\mathcal{H},  \label{Ham1}
\end{eqnarray}%
\begin{eqnarray}
\mathbf{\dot{\mu}} &\mathbf{=}&-\partial _{\mathbf{w}}\mathcal{H}%
+Div\partial _{\nabla \mathbf{w}}\mathcal{H},  \notag \\
\mathbf{\dot{w}} &\mathbf{=}&\partial _{\mathbf{\mu }}\mathcal{H},
\label{Ham2}
\end{eqnarray}%
which are Hamilton equations for IIC. In the case of IQ, $\mathbf{w}$
disappears in the list of entries of $\mathcal{H}$ and (\ref{Ham2}) reduces
to%
\begin{eqnarray}
\mathbf{\dot{\mu}} &=&Div\partial _{\nabla \mathbf{w}}\mathcal{H},  \notag \\
\mathbf{\dot{w}} &\mathbf{=}&\partial _{\mathbf{\mu }}\mathcal{H}.
\end{eqnarray}

Both in the case of IIC and IQ, general boundary conditions of the type%
\begin{equation}
\mathbf{x}\left( \mathbf{X}\right) =\mathbf{\bar{x}}\text{ \ \ \ \ \ on }%
\partial ^{\left( \mathbf{x}\right) }\mathcal{B}_{0},  \label{Bou1}
\end{equation}%
\begin{equation}
\partial _{\mathbf{F}}\mathcal{H}\mathbf{n=t}\text{ \ \ \ \ \ \ \ \ \ \ on }%
\partial ^{\left( \mathbf{t}\right) }\mathcal{B}_{0},
\end{equation}%
\begin{equation}
\mathbf{w}\left( \mathbf{X}\right) =\mathbf{\bar{w}}\text{ \ \ \ \ \ \ on }%
\partial ^{\left( \mathbf{w}\right) }\mathcal{B}_{0},  \label{Bou2}
\end{equation}%
\begin{equation}
\partial _{\nabla \mathbf{w}}\mathcal{H}\mathbf{n=}\mathfrak{t}\text{ \ \ \
\ \ \ \ \ \ \ \ on }\partial ^{\left( \mathfrak{t}\right) }\mathcal{B}_{0};
\label{Trac}
\end{equation}%
hold, where $\mathbf{\bar{x}}$, $\mathbf{t}$, $\mathbf{\bar{w}}$\ and $%
\mathfrak{t}$\ are prescribed on the relevant parts $\partial ^{\left( 
\mathbf{\cdot }\right) }\mathcal{B}_{0}$ of the boundary, chosen to be such
that $\partial ^{\left( \mathbf{x}\right) }\mathcal{B}_{0}\cap \partial
^{\left( \mathbf{t}\right) }\mathcal{B}_{0}=\varnothing $ with $Cl\left(
\partial \mathcal{B}_{0}\right) =Cl\left( \partial ^{\left( \mathbf{x}%
\right) }\mathcal{B}_{0}\cup \partial ^{\left( \mathbf{t}\right) }\mathcal{B}%
_{0}\right) $, and $\partial ^{\left( \mathbf{w}\right) }\mathcal{B}_{0}\cap
\partial ^{\left( \mathfrak{t}\right) }\mathcal{B}_{0}=\varnothing $ with $%
Cl\left( \partial \mathcal{B}_{0}\right) =Cl\left( \partial ^{\left( \mathbf{%
\nu }\right) }\mathcal{B}_{0}\cup \partial ^{\left( \mathfrak{t}\right) }%
\mathcal{B}_{0}\right) $, where $Cl$ indicates closure and $\mathbf{n}$ is
the outward unit normal to $\partial \mathcal{B}_{0}$ at all points in which
it is well defined.

Hamilton equations above are special cases of the ones discussed in [CM],
where general order parameter fields taking values on an abstract manifold
are accounted for.

It is rather difficult to imagine a loading device prescribing phason
tractions $\mathfrak{t}$ at the boundary. Problems with traction data might
thus involve the existence of at least one surface density $U\left( \mathbf{w%
}\right) $ such that $\mathfrak{t}=\rho _{0}\partial _{\mathbf{w}}U$ if not
another density of the type $\bar{U}\left( \mathbf{x}\right) $\ with $%
\mathbf{t}=\rho _{0}\partial _{\mathbf{x}}\bar{U}$. In this way one
considers the external boundary as a structured surface enveloping the body.

In this case, the Hamiltonian $H$ of the whole body is then given by%
\begin{equation*}
H\left( \mathbf{x,p,w,\mu }\right) =\int_{\mathcal{B}_{0}}\mathcal{H}\left( 
\mathbf{X,x,p,w,\mu }\right) d\left( vol\right) -
\end{equation*}%
\begin{equation}
-\int_{\partial ^{\left( 2\right) }\mathcal{B}_{0}}\left( \bar{U}\left( 
\mathbf{x}\right) -U\left( \mathbf{w}\right) \right) d\left( area\right) .
\end{equation}%
where $\partial ^{\left( 2\right) }\mathcal{B}_{0}=\partial ^{\left( \mathbf{%
t}\right) }\mathcal{B}_{0}\cup \partial ^{\left( \mathfrak{t}\right) }%
\mathcal{B}_{0}$. Notice that we write $\mathcal{H}\left( \mathbf{%
X,x,p,w,\mu }\right) $ instead of $\mathcal{H}\left( \mathbf{X,x,p,F,w,\mu ,}%
\nabla \mathbf{w}\right) $ because below we consider directly variational
derivatives.

\textbf{Theorem 4}. \emph{The canonical Hamilton equation}%
\begin{equation}
\dot{F}=\left\{ F,H\right\}  \label{Poisson}
\end{equation}%
\emph{is equivalent to the Hamiltonian system of balance equations }(\ref%
{Ham1})-(\ref{Ham2})\emph{\ for a quasiperiodic body where F is any
functional of the type }$\int_{\mathcal{B}_{0}}f\left( \mathbf{X,x,p,w,\mu }%
\right) $\emph{, with f a sufficiently smooth scalar density, and the
bracket }$\left\{ \cdot ,\cdot \right\} $\emph{\ is given by}%
\begin{eqnarray}
\left\{ F,H\right\} &=&\int_{\mathcal{B}_{0}}\left( \frac{\delta f}{\delta 
\mathbf{x}}\cdot \frac{\delta \mathcal{H}}{\delta \mathbf{p}}-\frac{\delta 
\mathcal{H}}{\delta \mathbf{x}}\cdot \frac{\delta f}{\delta \mathbf{p}}%
\right) d\left( vol\right) +  \notag \\
&&+\int_{\partial ^{\left( \mathbf{t}\right) }\mathcal{B}_{0}}\left( \frac{%
\delta f}{\delta \mathbf{x}}\cdot \frac{\delta \mathcal{H}}{\delta \mathbf{p}%
}\left\vert _{\partial ^{\left( \mathbf{t}\right) }\mathcal{B}_{0}}\right. -%
\frac{\delta \mathcal{H}}{\delta \mathbf{x}}\cdot \frac{\delta f}{\delta 
\mathbf{p}}\left\vert _{\partial ^{\left( \mathbf{t}\right) }\mathcal{B}%
_{0}}\right. \right) d\left( area\right) +  \notag \\
&&+\int_{\partial ^{\left( \mathfrak{t}\right) }\mathcal{B}_{0}}\left( \frac{%
\delta f}{\delta \mathbf{w}}\cdot \frac{\delta \mathcal{H}}{\delta \mathbf{%
\mu }}\left\vert _{\partial ^{\left( \mathfrak{t}\right) }\mathcal{B}%
_{0}}\right. -\frac{\delta \mathcal{H}}{\delta \mathbf{w}}\cdot \frac{\delta
f}{\delta \mathbf{\mu }}\left\vert _{\partial ^{\left( \mathfrak{t}\right) }%
\mathcal{B}_{0}}\right. \right) d\left( area\right) +  \notag \\
&&+\int_{\mathcal{B}_{0}}\left( \frac{\delta f}{\delta \mathbf{w}}\cdot 
\frac{\delta \mathcal{H}}{\delta \mathbf{\mu }}-\frac{\delta \mathcal{H}}{%
\delta \mathbf{\mu }}\cdot \frac{\delta f}{\delta \mathbf{w}}\right) d\left(
vol\right) ,  \label{Bracket}
\end{eqnarray}%
\emph{where the variational derivative} $\frac{\delta \mathcal{H}}{\delta 
\mathbf{x}}$ \emph{is obtained fixing }$\mathbf{p}$\emph{\ and allowing }$%
\mathbf{x}$\emph{\ to vary; an analogous meaning is valid for the
variational derivative with respect to the phason degree of freedom.}

The proof follows by direct calculation (see a more general version of this
theorem in [CM]).

\textbf{Remark 4}. The bracket $\left\{ \cdot ,\cdot \right\} $ is bilinear,
skew symmetric and satisfies Jacobi identity.

\textbf{Remark 5}. For $F=H$, (\ref{Poisson}) coincides with the equation of
conservation of energy.

\ \ \ \ \ \ \ \ \ \ \ \ 

Let us consider a boundary value problem in which traction data are not
prescribed at the boundary where just conditions like (\ref{Bou1}) and (\ref%
{Bou2}) hold. In this case (\ref{Bracket}) reduces to%
\begin{equation}
\left\{ F,H\right\} =\int_{\mathcal{B}_{0}}\left\{ f,\mathcal{H}\right\} _{P}%
\text{ }d\left( vol\right) ,
\end{equation}%
where%
\begin{equation}
\left\{ f,\mathcal{H}\right\} _{P}=\left( \frac{\delta f}{\delta \mathbf{x}}%
\cdot \frac{\delta \mathcal{H}}{\delta \mathbf{p}}-\frac{\delta \mathcal{H}}{%
\delta \mathbf{x}}\cdot \frac{\delta f}{\delta \mathbf{p}}\right) +\left( 
\frac{\delta f}{\delta \mathbf{w}}\cdot \frac{\delta \mathcal{H}}{\delta 
\mathbf{\mu }}-\frac{\delta \mathcal{H}}{\delta \mathbf{w}}\cdot \frac{%
\delta f}{\delta \mathbf{\mu }}\right) .
\end{equation}

It is matter of simple calculation to verify that $\left\{ \cdot ,\mathcal{%
\cdot }\right\} _{P}$ is bilinear, skew-symmetric, satisfies Jacobi identity
and also Leibniz identity. Then, $\left\{ \cdot ,\mathcal{\cdot }\right\}
_{P}$ induces relevant Poisson structures.

A purely spatial representation of all structures above described is
available with all fields defined over $\mathcal{B}$ rather than $\mathcal{B}%
_{0}$. Then $\mathcal{B}_{0}$ does not appear more neither as a reference
place nor as a model of paragon for quantities involved in the mechanical
model. In a spatial representation we start assuming a structure for
Hamiltonian density of the form%
\begin{equation}
\mathcal{\tilde{H}}\left( \mathbf{x,p,g,w}_{a}\mathbf{,\mu }_{a}\mathbf{,}%
grad\mathbf{w}_{a}\right) ,
\end{equation}%
where now, at a given $t$, $\mathbf{\mu }_{a}$\ is the canonical momentum
conjugated with $\mathbf{\dot{w}}_{a}$ and $\mathcal{H}$ depends also on the
spatial metric $\mathbf{g}$ rather than the gradient of deformation $\mathbf{%
F}$ because no reference to $\mathcal{B}_{0}$\ is made. In this case,
Hamilton equations read for IIC%
\begin{eqnarray}
\mathbf{\dot{p}} &=&-\partial _{\mathbf{\dot{x}}}\mathcal{H}+div\left(
2\partial _{\mathbf{g}}\mathcal{H}-\left( grad\mathbf{w}\right) ^{T}\partial
_{grad\mathbf{w}}\mathcal{H}\right) ,  \notag \\
\mathbf{\dot{x}} &=&\partial _{\mathbf{p}}\mathcal{H},
\end{eqnarray}%
\begin{eqnarray}
\mathbf{\dot{\mu}} &=&-\partial _{\mathbf{\dot{w}}}\mathcal{H}+div\partial
_{grad\mathbf{w}}\mathcal{H},  \notag \\
\mathbf{\dot{w}} &=&\partial _{\mathbf{\mu }}\mathcal{H},
\end{eqnarray}%
with the obvious reduction to IQ. Here, the Cauchy stress $\mathbf{\sigma }$
is then given by%
\begin{equation}
\mathbf{\sigma }=2\partial _{\mathbf{g}}\mathcal{H}-\left( grad\mathbf{w}%
\right) ^{T}\partial _{grad\mathbf{w}}\mathcal{H}.
\end{equation}

The term $\left( grad\mathbf{w}\right) ^{T}\partial _{grad\mathbf{w}}%
\mathcal{H}$\ rules an exchange of energy between the gross scale of
macroscopic deformation and the finer scale of phason changes, and vice
versa. An analogous phenomenon occurs in complex fluids where topological
transitions in the flows may be generated by this type of energy exchanges
[M03].

\section{Surfaces of discontinuity: conservative behavior}

Surfaces across which some quantities undergo finite jumps may occur in IIC
and IQ. They may be shock or acceleration waves, dislocations, closed cracks
and so on (see also remarks in [ML]). To describe their behavior, say rules
for their potential evolution, one needs not only to account for their
geometry and the action of standard interactions due to deformation
processes, but also to phason interactions.

Below we consider discontinuity surfaces endowed with own surface energy
which models the physical circumstance that interfaces are often regions in
a metastable state with an high concentration of energy [ML]. We allow not
only discontinuities of the standard gradient of deformation $\mathbf{F}$
across the surface but also jumps of $\mathbf{w}$ and its gradient. Really,
one may argue that the presence of $\nabla \mathbf{w}$ in the list of
entries of the energy takes into account in a regularized way the possible
presence of grain boundaries. This is true when the grain boundary is
between two regions with constant phason activity. However, in presence of
defects or in presence of subgrains containing `worms' (i.e. the topological
alterations of lattices assuring quasiperiodicity), we may have \emph{%
interaction between diffuse interfaces and sharp discontinuity surfaces}.
This is exactly the situation that we are analyzing here.

\subsection{A single discontinuity surface in $\mathcal{B}_{0}$}

Let $\Sigma $ be a single surface coinciding with $\left\{ \mathbf{X}\in cl%
\mathcal{B}_{0}\text{, \ }g\left( \mathbf{X}\right) =0\right\} $, where $g$
is a smooth function (smoothness chosen for the sake of simplicity) with
non-singular gradient. It is oriented by the normal vector field $\mathbf{m=%
\tilde{m}}\left( \mathbf{X}\right) =\nabla g\left( \mathbf{X}\right)
/\left\vert \nabla g\left( \mathbf{X}\right) \right\vert $.

Let $\mathbf{X}\longmapsto a=\tilde{a}\left( \mathbf{X}\right) $ be a field
taking values in a linear space and suffering bounded discontinuities across 
$\Sigma $. For $\varepsilon >0$ we indicate with $a^{\pm }$ the limits $\lim 
_{\substack{ \varepsilon \rightarrow 0  \\ \mathbf{X}\in \Sigma }}a\left( 
\mathbf{X}\pm \varepsilon \mathbf{m}\right) $. The jump $[a]$ of $a$ across $%
\Sigma $ is defined by $[a]=a^{+}-a^{-}$ while the average $\left\langle
a\right\rangle $\ by $2\left\langle a\right\rangle =a^{+}+a^{-}$. If fields $%
a_{1}$ and $a_{2}$ have the same properties of $a$ we have $%
[a_{1}a_{2}]=[a_{1}]\left\langle a_{2}\right\rangle +\left\langle
a_{1}\right\rangle [a_{2}]$ with the product $a_{1}a_{2}$\ defined in some
way assuring distributivity.

$\Sigma $ is \emph{coherent} when at each $\mathbf{X}$ one gets $[\mathbf{F}%
]\left( \mathbf{I-m\otimes m}\right) =\mathbf{0}$.

For any `virtual' motion to $\Sigma $ prescribed by means of a vector field%
\begin{equation}
\Sigma \ni \mathbf{X}\overset{\mathbf{\tilde{\varpi}}}{\mathbf{\longmapsto }}%
\mathfrak{\varpi }=\mathbf{\tilde{\varpi}}\left( \mathbf{X}\right) \in 
\mathbb{R}^{3}
\end{equation}%
with normal component $U=\mathbf{\varpi \cdot m}$ and assume that the
velocity $\mathbf{\dot{x}}$ may suffer bounded jumps across $\Sigma $, we
get the condition $\left[ \mathbf{\dot{x}}\right] =-U\left[ \mathbf{F}\right]
\mathbf{m}$.

At each $\mathbf{X}\in \Sigma $ we define the surface deformation gradient $%
\mathbb{F}$ as $\left\langle \mathbf{F}\right\rangle \left( \mathbf{%
I-m\otimes m}\right) \in Hom\left( T_{\mathbf{X}}\Sigma ,T_{\mathbf{x}}%
\mathcal{B}\right) $ and indicate with $\mathbb{N}$ the projection over $%
\Sigma $ of the average of $\nabla \mathbf{w}$, namely $\left\langle \nabla 
\mathbf{w}\right\rangle \left( \mathbf{I-m\otimes m}\right) \in Hom\left( T_{%
\mathbf{X}}\Sigma ,T_{\mathbf{w}}\mathbb{V}_{w}\right) $.

\subsection{Phonon and phason surface measures of interaction and their
balance}

We consider $\Sigma $ endowed with a \emph{surface energy density} $\phi $
assumed to be sufficiently smooth and given by%
\begin{equation}
\left( \mathbf{m,}\mathbb{F}\mathbf{,}\left\langle \mathbf{w}\right\rangle 
\mathbf{,}\mathbb{N}\right) \overset{\tilde{\phi}}{\longmapsto }\phi =\tilde{%
\phi}\left( \mathbf{m,}\mathbb{F}\mathbf{,\left\langle \mathbf{w}%
\right\rangle ,}\mathbb{N}\right) .
\end{equation}%
The dependence of $\phi $\ on $\mathbf{m}$ accounts for \emph{anisotropy} of 
$\Sigma $.

We require the invariance of $\phi $ with respect to

\begin{description}
\item[(i)] changes of observers and

\item[(ii)] relabeling of $\Sigma $.
\end{description}

As discussed above, changes of observers are characterized by the action of
the group of automorphisms of $\mathcal{E}^{3}$ and of a generic Lie group
over $\mathbb{V}_{w}$. However, the definition of $\mathbf{f}_{s_{1}}^{1}$
needs to be modified in order to describe the relabeling of $\Sigma $ in
addition to the overall relabeling of $\mathcal{B}_{0}$.

We should then consider time-parametrized families $s_{1}\longmapsto \mathbf{%
\hat{f}}_{s_{1}}^{1}$\ of elements of $SDiff\left( \mathcal{B}_{0}\right) $
characterized by the properties listed below (see [dFM]).

\begin{enumerate}
\item The map $s_{1}\longmapsto \mathbf{\hat{f}}_{s_{1}}^{1}$ satisfies A1.
Moreover, the field $\mathcal{B}_{0}\ni \mathbf{X}\longmapsto \mathfrak{w}=%
\mathfrak{\tilde{w}}\left( \mathbf{X}\right) =\mathbf{\hat{f}}%
_{0}^{1^{\prime }}\left( \mathbf{X}\right) $ is at least of class $%
C^{1}\left( \mathcal{B}_{0}\right) $, then across and along $\Sigma $.

\item Each $\mathbf{\hat{f}}_{s_{1}}^{1}$ preserves the elements of area of $%
\Sigma $. Namely, if $dA$ is the element of area of $\Sigma $ in $\mathcal{B}%
_{0}$, $dA=\mathbf{\hat{f}}_{s_{1}}^{1\ast }\circ dA$, where the asterisk
indicates push forward.

\item $\left( \nabla \mathfrak{w}\right) \mathbf{m=0}$.

\item $\nabla _{\Sigma }\mathsf{v}_{m}=0$, with $\mathsf{v}_{m}=\mathfrak{w}%
\cdot \mathbf{m}$.
\end{enumerate}

\ \ \ \ \ \ \ \ \ \ \ \ \ \ \ \ \ \ \ 

\textbf{Definition 2} (invariance of $\phi $). A surface energy density $%
\phi $ is invariant with respect to the action of $\mathbf{\hat{f}}%
_{s_{1}}^{1}$, $\mathbf{f}_{s_{2}}^{2}$ and $G$ if%
\begin{equation}
\tilde{\phi}\left( \mathbf{m,}\mathbb{F}\mathbf{,\left\langle \mathbf{w}%
\right\rangle ,}\mathbb{N}\right) =\tilde{\phi}\left( \nabla \mathbf{\hat{f}}%
^{1T}\mathbf{m,}\left( grad_{\Sigma }\mathbf{f}^{2}\right) \mathbb{F}\left(
\nabla \mathbf{\hat{f}}^{1}\right) ^{-1}\mathbf{,\left\langle \mathbf{w}%
\right\rangle }_{g}\mathbf{,}\mathbb{N}_{g}\left( \nabla \mathbf{\hat{f}}%
^{1}\right) ^{-1}\right) ,
\end{equation}%
for any $g\in G$ and $s_{1},s_{2}\in \mathbb{R}^{+}$, where $\mathbb{N}%
_{g}=\left\langle \nabla \mathbf{w}_{g}\right\rangle \left( \mathbf{%
I-m\otimes m}\right) $ and we have used notations common to Definition 1.

\ \ \ \ \ \ \ \ \ \ \ \ \ \ \ \ \ \ \ \ \ \ \ \ \ 

Let $\mathfrak{X}$ be a sufficiently smooth vector density defined over $%
\Sigma $ by%
\begin{equation}
\mathfrak{X}=-\phi \Pi \mathbf{w}+\left( \partial _{\mathbb{F}}\phi \right)
^{T}\left( \mathbf{v}-\left\langle \mathbf{F}\right\rangle \mathfrak{w}%
\right) +\left( \partial _{\mathbb{N}}\phi \right) ^{T}\left( \xi _{\mathbb{V%
}_{\mathbf{w}}}\left( \mathbf{\left\langle \mathbf{w}\right\rangle }\right)
-\left\langle \nabla \mathbf{w}\right\rangle \mathfrak{w}\right) -\left(
\partial _{\mathbf{m}}\phi \otimes \mathbf{m}\right) \mathfrak{w}\mathbf{.}
\end{equation}

\ \ \ \ \ \ \ \ \ \ \ \ \ \ \ \ \ \ \ \ \ \ \ 

\textbf{Theorem 5}. \emph{Let} $\Sigma $ \emph{be a structured surface with
surface energy} $\phi $. \emph{Let us assume}%
\begin{equation}
\frac{d}{dt}\int_{\mathfrak{b}_{\Sigma }}\mathcal{Q}d\left( vol\right)
+\int_{\partial \mathfrak{b}_{\Sigma }}\mathfrak{F}\cdot \mathbf{n}d\left(
area\right) +\int_{\partial \left( \mathfrak{b}_{\Sigma }\cap \Sigma \right)
}\mathfrak{X}\cdot \mathsf{n}d\left( length\right) =0  \label{IntSuf}
\end{equation}%
\emph{for any part} $\mathfrak{b}_{\Sigma }$ \emph{of} $\mathcal{B}_{0}$ 
\emph{crossing} $\Sigma $. \emph{If} $\mathcal{L}$ \emph{and} $\phi $\ \emph{%
are invariant with respect to} $\mathbf{\hat{f}}_{s_{1}}^{1}$, $\mathbf{f}%
_{s_{2}}^{2}$ \emph{and} $G$\emph{, covariant pointwise balances across} $%
\Sigma $ \emph{follow as in the list below.}

\begin{enumerate}
\item \emph{The action of }$\mathbf{f}_{s_{2}}^{2}$ \emph{alone implies the
interfacial balance of standard interactions}%
\begin{equation}
\left[ \mathbf{P}\right] \mathbf{m}+Div_{\Sigma }\mathbb{T}=-\rho _{0}\left[ 
\mathbf{\dot{x}}\right] U,  \label{GS}
\end{equation}%
\emph{where} $\mathbb{T=-}\partial _{\mathbb{F}}\phi \in Hom\left( T_{%
\mathbf{X}}\Sigma ,T_{\mathbf{x}}^{\ast }\mathcal{B}\right) $ \emph{is the
surface Piola-Kirchhoff stress}.

\item \emph{The action of} $G$ \emph{alone implies the interfacial balance
of substructural interactions}%
\begin{equation}
\left[ \mathcal{S}\right] \mathbf{m}+Div_{\Sigma }\mathbb{S}-\mathfrak{z}=-%
\bar{\rho}\left[ \mathbf{\dot{w}}\right] U,  \label{PMM1}
\end{equation}%
\emph{where }$\mathbb{S=-}\partial _{\mathbb{N}}\phi \in Hom\left( T_{%
\mathbf{X}}\Sigma ,T_{\mathbf{w}}^{\ast }\mathbb{V}_{w}\right) $ \emph{is
the surface microstress and} $\mathfrak{z}=\partial _{\mathbf{w}}\phi \in T_{%
\mathbf{w}}^{\ast }\mathbb{V}_{w}$ \emph{the surface self-force}.

\item \emph{The action of} $\mathbf{\hat{f}}_{s_{1}}^{1}$ \emph{alone
implies the interfacial configurational balance along the normal} $\mathbf{m}
$ \emph{in absence of dissipative forces driving} $\Sigma $\emph{, namely}%
\begin{equation*}
\mathbf{m\cdot }\left[ \mathbb{P}\right] \mathbf{m}+\mathbb{C}_{\tan }\cdot 
\mathsf{L}+Div_{\Sigma }\mathfrak{c}=
\end{equation*}%
\begin{equation}
=\bar{\rho}U\left[ \left( \nabla \mathbf{w}\right) ^{T}\mathbf{\dot{w}}%
\right] \cdot \mathbf{m}+\frac{1}{2}\bar{\rho}\left[ \left\vert \mathbf{\dot{%
w}}\right\vert ^{2}\right] -\frac{1}{2}\rho _{0}U^{2}\left[ \left\vert 
\mathbf{Fm}\right\vert ^{2}\right] ,  \label{PMM2}
\end{equation}%
\emph{where}%
\begin{equation}
\mathbb{C}_{\tan }=\phi \Pi -\mathbb{F}^{T}\mathbb{T-N}^{T}\mathbb{S}
\label{66}
\end{equation}%
\emph{is a generalized version of the surface Eshelby stress and}%
\begin{equation}
\mathfrak{c}=-\partial _{\mathbf{m}}\phi -\mathbb{T}^{T}\left\langle \mathbf{%
F}\right\rangle \mathbf{m}-\mathbb{S}^{T}\left\langle \nabla \mathbf{w}%
\right\rangle \mathbf{m}  \label{67}
\end{equation}%
\emph{is a surface shear}.
\end{enumerate}

\ \ \ \ \ \ \ \ \ \ \ \ \ \ \ \ \ \ \ \ \ 

An analogous theorem can be found in [dFM]. However, though there order
parameters taking values in an abstract manifold $\mathcal{M}$\ are
considered instead of phonon modes (so, there, the point of view involves a
unifying framework for models of condensed matter physics), such order
parameters are assumed to be continuous across $\Sigma $. Here, on the
contrary, we allow jumps of $\mathbf{w}$. The proof below follows the one of
the theorem in [dFM] quoted above. We adapt it to the situation envisaged
here with slight modifications and report it with a certain number of
details just for the sake of completeness. In any case when in multifield
theories the order parameter field takes values on a linear space, it can be
considered discontinuous at the sharp discontinuity surface (if it exists)
and the counterpart of theorem above holds. On the contrary, when the
manifold $\mathcal{M}$ of substructural morphologies does not coincide with
a linear space, since $\mathcal{M}$ has finite dimension, it can be embedded
isometrically in an appropriate linear space. However, the embedding itself
becomes a prominent part of modeling. Although the isometric embedding is
preferable because it preserves the quadratic part of the substructural
kinetic energy (if it exists as in IIC), in fact, such an embedding is not
unique (as non-isometric ones) and also not `rigid'. Then, the selection of
the appropriate embedding (if necessary in the case of abstract order
parameters) is not simple and general criteria suggested by physical
instances seems to be not known.

\ \ \ \ \ \ \ \ \ \ \ \ \ \ \ \ \ \ 

\textbf{Proof.}

Conditions assuring the invariance of $\tilde{\phi}$\ with respect to
changes of observers and relabeling are given by $\frac{d}{ds_{i}}\phi
\left\vert _{s_{1}=0,s_{2}=0,s_{3}=0}\right. =0$, with $i=1,2,3$. They
correspond to%
\begin{equation}
\mathbb{F}^{T}\mathbb{T\cdot \nabla }_{\Sigma }\mathfrak{w}+\mathbb{N}^{T}%
\mathbb{S\cdot \nabla }_{\Sigma }\mathfrak{w}\mathbf{+\partial }_{\mathbf{m}%
}\phi \cdot \left( \nabla \mathbf{w}\right) \mathfrak{w}=0,  \label{NR1}
\end{equation}%
\begin{equation}
\mathbb{T\cdot \nabla }_{\Sigma }\mathbf{v}=0,  \label{NR2}
\end{equation}%
\begin{equation}
\mathfrak{z}\cdot \xi _{\mathbb{V}_{\mathbf{w}}}\left( \mathbf{\left\langle 
\mathbf{w}\right\rangle }\right) +\mathbb{S\cdot \nabla }_{\Sigma }\xi _{%
\mathbb{V}_{\mathbf{w}}}\left( \mathbf{\left\langle \mathbf{w}\right\rangle }%
\right) =0.  \label{NR3}
\end{equation}

If we shrink $\mathfrak{b}_{\Sigma }$ to $\mathfrak{b}_{\Sigma }\cap \Sigma $
uniformly in time, we get the pointwise balance (see [dFM])%
\begin{equation}
-[\mathcal{Q]}U+[\mathfrak{F]}\cdot \mathbf{m}+Div_{\Sigma }\mathfrak{X}=0,
\label{IntNet}
\end{equation}%
as a consequence of the arbitrariness of $\mathfrak{b}_{\Sigma }$.

If $\mathbf{f}^{2}$ acts alone, then%
\begin{equation}
\mathfrak{X}=\mathbb{T}^{T}\mathbf{v},\text{ \ \ }\mathcal{Q}=\rho \mathbf{%
\dot{x}\cdot v},\text{ \ \ }\mathfrak{F}=-\mathbf{P}^{T}\mathbf{v,}
\end{equation}%
so that, as a consequence of (\ref{NR2}), $Div_{\Sigma }\mathfrak{X}=\mathbf{%
v\cdot }Div_{\Sigma }\mathbb{T}.$ The arbitrariness of $\mathbf{v}$ and its
continuity across $\Sigma $ implies (\ref{GS}) from (\ref{IntNet}).

If $G$ acts alone, we get%
\begin{equation}
\mathfrak{X}=\mathbb{S}^{T}\xi _{\mathbb{V}_{\mathbf{w}}}\left( \mathbf{%
\left\langle \mathbf{w}\right\rangle }\right) ,\text{ \ \ }\mathcal{Q}=\bar{%
\rho}\mathbf{\dot{w}\cdot }\xi _{\mathbb{V}_{\mathbf{w}}}\left( \mathbf{%
\left\langle \mathbf{w}\right\rangle }\right) ,\text{ \ \ }\mathfrak{F}=-%
\mathcal{S}^{T}\xi _{\mathbb{V}_{\mathbf{w}}}\left( \mathbf{\left\langle 
\mathbf{w}\right\rangle }\right) \mathbf{,}
\end{equation}%
and, from (\ref{NR3}),%
\begin{equation}
Div_{\Sigma }\mathfrak{X}=\xi _{\mathbb{V}_{\mathbf{w}}}\left( \mathbf{%
\left\langle \mathbf{w}\right\rangle }\right) \cdot \left( Div_{\Sigma }%
\mathbb{S}-\mathfrak{z}\right) .
\end{equation}%
Then, from (\ref{IntNet}) we obtain (\ref{PMM1}) thanks to the arbitrariness
of the element $\xi $\ selected in the Lie algebra of $G$.

If $\mathbf{\hat{f}}^{1}$ acts alone, then%
\begin{equation}
\mathcal{Q}=-\rho _{0}\mathbf{F}^{T}\mathbf{\dot{x}}\cdot \mathfrak{w}%
\mathbf{-}\bar{\rho}\left( \nabla \mathbf{w}\right) ^{T}\mathbf{\dot{w}}%
\cdot \mathfrak{w}\mathbf{,}
\end{equation}%
\begin{equation}
\mathfrak{F}=\left( \left( \frac{1}{2}\rho _{0}\left\vert \mathbf{\dot{x}}%
\right\vert ^{2}+\frac{1}{2}\bar{\rho}\left\vert \mathbf{\dot{w}}\right\vert
^{2}\right) \mathbf{I-}\mathbb{P}\right) \mathfrak{w}
\end{equation}%
\begin{equation}
\mathfrak{X}=-\mathbb{C}_{\tan }^{T}\mathfrak{w}\mathbf{-}\mathfrak{c}%
\mathsf{v}_{m},
\end{equation}%
with $\mathbb{C}_{\tan }$\ and $\mathfrak{c}$\ defined respectively by (\ref%
{66}) and (\ref{67}) and $\mathsf{v}_{m}=\mathfrak{w}\cdot \mathbf{m}$.

Terms of equation (\ref{IntNet}) then become in this case%
\begin{equation*}
-[\mathcal{Q]}U+[\mathfrak{F]}\cdot \mathbf{m}=\rho _{0}[\mathbf{F}^{T}%
\mathbf{\dot{x}}]U\cdot \mathfrak{w}\mathbf{+}\bar{\rho}[\left( \nabla 
\mathbf{w}\right) ^{T}\mathbf{\dot{w}}]U\cdot \mathfrak{w}\mathbf{+}
\end{equation*}%
\begin{equation}
+\frac{1}{2}\rho _{0}[\left\vert \mathbf{\dot{x}}\right\vert ^{2}]\mathfrak{w%
}\cdot \mathbf{m}+\frac{1}{2}\bar{\rho}[\left\vert \mathbf{\dot{w}}%
\right\vert ^{2}]\mathfrak{w}\cdot \mathbf{m-[}\mathbb{P}\mathbf{]}\mathfrak{%
w}\cdot \mathbf{m,}  \label{C}
\end{equation}%
\begin{equation}
Div_{\Sigma }\left( \mathbb{C}_{\tan }^{T}\mathfrak{w}+\mathfrak{c}\mathsf{v}%
_{m}\right) =\mathfrak{w}\cdot \left( Div_{\Sigma }\mathbb{C}_{\tan }+\left(
Div_{\Sigma }\mathfrak{c}\right) \mathbf{m}\right) ,  \label{DivC}
\end{equation}%
where the second equation is a consequence of (\ref{NR1}), the circumstance
that $\left( \mathbf{I-m\otimes m}\right) \cdot \nabla _{\Sigma }\mathfrak{w}%
=\left( \left( \nabla \mathfrak{w}\right) \mathbf{m}\right) \cdot \mathbf{m}$%
,\ since $\mathfrak{w}$\ is isocoric, and properties 3 and 4 of the
definition of the relabeling $\mathbf{\hat{f}}^{1}$ of $\mathcal{B}_{0}$\
including $\Sigma $.

By inserting (\ref{C}) and (\ref{DivC}) in (\ref{IntNet}), the arbitrariness
of $\mathfrak{w}$ implies%
\begin{equation*}
\rho _{0}[\mathbf{F}^{T}\mathbf{\dot{x}}]U+\bar{\rho}[\left( \nabla \mathbf{w%
}\right) ^{T}\mathbf{\dot{w}}]U+\frac{1}{2}\rho _{0}[\left\vert \mathbf{\dot{%
x}}\right\vert ^{2}]\mathbf{m+}
\end{equation*}%
\begin{equation}
+\frac{1}{2}\bar{\rho}[\left\vert \mathbf{\dot{w}}\right\vert ^{2}]\mathbf{m}%
=\mathbf{[}\mathbb{P}^{T}\mathbf{]m}+Div_{\Sigma }\mathbb{C}_{\tan }+\left(
Div_{\Sigma }\mathfrak{c}\right) \mathbf{m}  \label{SurfC}
\end{equation}%
and we shall evaluate the component along $\mathbf{m}$ of (\ref{SurfC}).

By indicating with $\mathbf{\bar{v}}$ the averaged velocity $\mathbf{\bar{v}}%
=\left\langle \mathbf{\dot{x}}\right\rangle +U\left\langle \mathbf{F}%
\right\rangle \mathbf{m}$ and using the relation $[\mathbf{\dot{x}}]=-U[%
\mathbf{F}]\mathbf{m}$ mentioned previously, we then get%
\begin{equation}
\rho _{0}[\mathbf{F}^{T}\mathbf{\dot{x}}]U\cdot \mathbf{m}=-\frac{1}{2}[\rho
_{0}\left\vert \mathbf{\dot{x}-\bar{v}}\right\vert ^{2}]=\rho _{0}[\mathbf{%
\dot{x}}]\cdot \mathbf{\bar{v}}-\rho _{0}\mathbf{[}\left\vert \mathbf{\dot{x}%
}\right\vert ^{2}\mathbf{];}  \label{Box}
\end{equation}%
where $\frac{1}{2}[\rho _{0}\left\vert \mathbf{\dot{x}-\bar{v}}\right\vert
^{2}$ is the relative kinetic energy referred to $\Sigma $. We also get%
\begin{equation}
\frac{1}{2}\rho _{0}[\left\vert \mathbf{\dot{x}}\right\vert ^{2}]=-\rho _{0}[%
\mathbf{\dot{x}}]\cdot \mathbf{\bar{v}+}\frac{1}{2}\rho _{0}U^{2}[\left\vert 
\mathbf{Fm}\right\vert ^{2}].  \label{GiumChin}
\end{equation}%
by using once more $[\mathbf{\dot{x}}]=-U[\mathbf{F}]\mathbf{m}$ and the
definition of $\mathbf{\bar{v}}$. By evaluating the normal component of (\ref%
{SurfC}), using (\ref{Box}), (\ref{GiumChin}) and taking into account that $%
\mathbf{m\cdot }Div_{\Sigma }\mathbb{C}_{\tan }=\mathbb{C}_{\tan }\cdot 
\mathsf{L}$, as it is simple to verify (see Lemma 2 in [dFM]), we get (\ref%
{PMM2}) and the theorem is proven.

\textbf{Remark 6}. Of course, for unstructured interfaces, i.e. in absence
of surface energy, interfacial balances at items 1, 2 and 3 of Theorem 5
become respectively%
\begin{equation}
\left[ \mathbf{P}\right] \mathbf{m}=-\rho _{0}\left[ \mathbf{\dot{x}}\right]
U,
\end{equation}%
\begin{equation}
\left[ \mathcal{S}\right] \mathbf{m}=-\bar{\rho}\left[ \mathbf{\dot{w}}%
\right] U,
\end{equation}%
\begin{equation}
\mathbf{m\cdot }\left[ \mathbb{P}\right] \mathbf{m=}\bar{\rho}U\left[ \left(
\nabla \mathbf{w}\right) ^{T}\mathbf{\dot{w}}\right] \cdot \mathbf{m}+\frac{1%
}{2}\bar{\rho}\left[ \left\vert \mathbf{\dot{w}}\right\vert ^{2}\right] -%
\frac{1}{2}\rho _{0}U^{2}\left[ \left\vert \mathbf{Fm}\right\vert ^{2}\right]
.
\end{equation}

\section{Phason friction}

Non-conservative phenomena may occur in quasicrystals and involve different
mechanisms such as (for example) viscous effects or plastic flows. Below we
discuss just possible viscous effects due to `internal' friction of phason
nature. To fix ideas we restrict ourselves first to the case in which just
friction of purely local nature exists. To account for it we may follow
different ways. A classical one (see [BKMR] and [MR] for deep remarks about
it) involves an integral Lagrange-d'Alambert principle. Preliminarily, we
recall that balance equations (\ref{A}) and (\ref{B}) come from a
variational principle of the form%
\begin{equation}
\delta \tint\limits_{\mathcal{B}_{0}\times \lbrack 0,\bar{t}]}\mathcal{L}%
\left( j^{1}\left( \eta \right) \left( \mathbf{X},t\right) \right) \text{ }%
d\left( vol\right) \wedge dt=0.
\end{equation}%
To account for possible internal friction of pure phason nature, we may then
consider the following Lagrange-d'Alembert principle:%
\begin{equation}
\delta \left( \tint\limits_{\mathcal{B}_{0}\times \lbrack 0,\bar{t}]}%
\mathcal{L}\left( j^{1}\left( \eta \right) \left( \mathbf{X},t\right)
\right) \text{ }d\left( vol\right) \wedge dt\right) +\tint\limits_{\mathcal{B%
}_{0}\times \lbrack 0,\bar{t}]}\mathbf{z}^{v}\cdot \delta \mathbf{w}\text{ }%
d\left( vol\right) \wedge dt=0,
\end{equation}%
with

\begin{description}
\item[(a)] $\mathbf{z}^{v}=\mathbf{\tilde{z}}^{v}\left( \mathbf{F,w,}\nabla 
\mathbf{w,\dot{w}}\right) \in T_{\mathbf{w}}^{\ast }\mathbb{V}_{w},$

\item[(b)] $\mathbf{z}^{v}\cdot \mathbf{\dot{w}}\geq 0,$ \ \ $\forall 
\mathbf{\dot{w}}$.
\end{description}

The property (b) declares that the `viscous' self-force $\mathbf{z}^{v}$\ is
purely dissipative. In other words, one may say that formally (as it will be
clarified further in next section) a balance of phason interactions of the
form%
\begin{equation}
\bar{\rho}\mathbf{\ddot{w}}=-\mathbf{z}+Div\mathcal{S}  \label{SI}
\end{equation}%
still holds in non-conservative case, but with $\mathbf{z}$, the self-force
of phason nature, admitting an additive decomposition of the form $\mathbf{%
z=z}^{eq}+\mathbf{z}^{v}$, with $\mathbf{z}^{eq}$ the part coming from
thermodynamic equilibrium as in Corollary 2 and $\mathbf{z}^{v}$ of purely
dissipative nature. The abuse of notation between (\ref{BSI}) and (\ref{SI})
is rather negligible because though the measures of interactions involved
(namely phason stresses and self-forces) are placed within different
thermodynamical settings, their nature is the same: they represent in fact
interactions between neighboring material elements (the contact interactions
represented by $\mathcal{S}$) and interactions occurring within each
material element.

The inequality in (b) is satisfied by an expression of the type%
\begin{equation}
\mathbf{z}^{v}=c\mathbf{\dot{w},}  \label{Star}
\end{equation}%
with $c=\tilde{c}\left( \mathbf{F,w,}\nabla \mathbf{w,\dot{w}}\right) $ and $%
\tilde{c}$\ a scalar definite positive function such that $\tilde{c}\left( 
\mathbf{F,w,}\nabla \mathbf{w,}0\right) =0$. In this case the balance of
phason interactions becomes%
\begin{equation}
\bar{\rho}\mathbf{\ddot{w}}=Div\left( \partial _{\nabla \mathbf{w}}e\right)
-\partial _{\mathbf{w}}e-c\mathbf{\dot{w},}
\end{equation}%
in the general case, while its reduced form for IQ is given by%
\begin{equation}
c\mathbf{\dot{w}}=Div\left( \partial _{\nabla \mathbf{w}}e\right) .
\label{Min}
\end{equation}%
Of course, property (b) may imply explicit forms of $\mathbf{z}^{v}$ more
articulated than (\ref{Star}) and involving a structure of the type $\mathbf{%
z}^{v}=\mathbf{A\dot{w}}$\ with $\mathbf{A}$ a second order definite
positive tensor. Constitutive structures for $\mathbf{z}^{v}$ involving
tensor coefficients may also satisfy frame indifferent conditions (see
discussions in [A] and [Si] about viscous stresses in classical
viscoelasticity).

Equation (\ref{Min}) fits the minimal model proposed in [RoLo].

We may consider also phason friction effects of weakly non-local (or better,
gradient) nature. To this end we may consider not only a thermodynamic non
equilibrium part $\mathbf{z}^{v}$ of the internal self-force appearing in
the sum $\mathbf{z=z}^{eq}+\mathbf{z}^{v}$ but also a dissipative phason
stress $\mathcal{S}^{v}$ satisfying the decomposition $\mathcal{S}=\mathcal{S%
}^{eq}+\mathcal{S}^{v}$ with $\mathcal{S}^{eq}$\ as in Corollary 2 and $%
\mathcal{S}^{v}$\ of purely dissipative nature expressed by the inequality%
\begin{equation}
\mathbf{z}^{v}\cdot \mathbf{\dot{w}}+\mathcal{S}^{v}\cdot \nabla \mathbf{%
\dot{w}}\geq 0  \label{Tri}
\end{equation}%
that we presume to be satisfied by any choice of the rates involved. A
possible solution of the inequality (\ref{Tri}) is given by%
\begin{equation}
\mathbf{z}^{v}=c^{\ast }\mathbf{\dot{w},}\text{ \ \ \ \ \ \ }\mathcal{S}%
^{v}=\omega \nabla \mathbf{\dot{w},}  \label{Rec}
\end{equation}%
with $c^{\ast }=\tilde{c}^{\ast }\left( \mathbf{F,w,}\nabla \mathbf{w,\dot{w}%
,}\nabla \mathbf{\dot{w}}\right) $, $\omega =\tilde{\omega}\left( \mathbf{%
F,w,}\nabla \mathbf{w,\dot{w},}\nabla \mathbf{\dot{w}}\right) $, and $\tilde{%
c}^{\ast }$, $\tilde{\omega}$\ scalar definite positive functions such that $%
\tilde{c}^{\ast }\left( \mathbf{F,w,}\nabla \mathbf{w,}0\mathbf{,}0\right)
=0 $ and $\tilde{\omega}\left( \mathbf{F,w,}\nabla \mathbf{w,}0\mathbf{,}%
0\right) =0$. Of course, (\ref{Rec}) is not the sole possible solution of (%
\ref{Tri}) because tensor coefficients may be involved and also linear
combinations of $\mathbf{\dot{w}}$\ and $\nabla \mathbf{\dot{w}}$\ (see for
a more general case [MA]). However, in the case of occurrence of (\ref{Tri}%
), the reduced balance (\ref{Min}) for IQ becomes%
\begin{equation}
c^{\ast }\mathbf{\dot{w}}=Div\left( \partial _{\nabla \mathbf{w}}e+\omega
\nabla \mathbf{\dot{w}}\right) .
\end{equation}

\section{$SO\left( 3\right) $ invariance and the nature of the balance of
phason interactions}

In deriving the balance (\ref{BSI}) we have mixed the representation of
phason interactions (obtained by means of the phason stress $\mathcal{S}$
and the phason self-force $\mathbf{z}$) and their constitutive structure
declared through the derivatives of the Lagrangian with respect to $\nabla 
\mathbf{w}$ and $\mathbf{w}$ respectively. However, we have also adopted the
formal counterpart of (\ref{BSI}), namely (\ref{SI}), in non-conservative
case, paying attention in the second circumstance to `viscous'-like parts of
the interactions. A question is whether such a pointwise balance holds
formally always before discussing constitutive issues. Another connected
question is whether an integral (global) version of the pointwise balance of
phason interactions can be postulated a-priori as a balance of
phason-momentum.

\begin{itemize}
\item The answer to the first question is affirmative: the balance of phason
interactions holds in the form (\ref{SI}) independently of constitutive
issues.

\item As regards the second question, though an integral version of (\ref{SI}%
) can be in principle postulated \emph{because} $\mathbb{V}_{w}$ is a linear
space, it is \emph{not} necessary because just the integral balance of
standard forces and a non-standard balance of couples suffices to get
pointwise balances.
\end{itemize}

To prove previous statements we leave constitutive issues out of
consideration and try to represent interactions just in their purely
geometric form as objects power conjugated with the rates of the descriptors
of the morphology of the body, namely phonon and phason degrees of freedom.

Let $\mathfrak{b}$ any part of $\mathcal{B}_{0}$, i.e. any subset of $%
\mathcal{B}_{0}$\ with non-null volume measure and the same regularity
properties of $\mathcal{B}_{0}$. We presume that the part in $\mathfrak{b}$\
interacts with the rest of the body and the external environment through
interactions of bulk and contact nature, the latter exerted through the
boundary $\partial \mathfrak{b}$. The external power $\mathcal{P}_{\mathfrak{%
b}}^{ext}\left( \mathbf{\dot{x},\dot{w}}\right) $ of all interactions over $%
\mathfrak{b}$, a linear functional over the space of rates $\mathbf{\dot{x}}$%
\ and $\mathbf{\dot{w}}$, is then given by%
\begin{equation}
\mathcal{P}_{\mathfrak{b}}^{ext}\left( \mathbf{\dot{x},\dot{w}}\right)
=\int_{\mathfrak{b}}\left( \mathbf{\bar{b}\cdot \dot{x}+\beta \cdot \dot{w}}%
\right) d\left( vol\right) +\int_{\partial \mathfrak{b}}\left( \mathbf{%
Pn\cdot \dot{x}}+\mathcal{S}\mathbf{n\cdot \dot{w}}\right) d\left(
area\right) .
\end{equation}%
Here, $\mathbf{\bar{b}}$ represents standard bulk forces and is decomposed
as $\mathbf{\bar{b}}=\rho _{0}\mathbf{b+b}^{in}$ where $\mathbf{b}$ is the
objective part which is coincident with the analogous $\mathbf{b}$ in
Corollary 1 while $\mathbf{b}^{in}$ is of pure inertial phonon nature. $%
\mathbf{\beta }$\ is of pure inertial phason nature (if phason inertia
exists) while $\mathbf{P}$ and $\mathcal{S}$ are respectively the first
Piola-Kirchhoff stress and the phason stress as in previous sections. $%
\mathbf{Pn}$\ represents the traction developing power in the relative
change of place of neighboring material elements at the boundary $\partial 
\mathfrak{b}$\ imagining the phason activity frozen. $\mathcal{S}\mathbf{n}$%
\ pictures interactions developed across the boundary $\partial \mathfrak{b}$%
\ between neighboring material elements which do not change place but
display different phason activity. As pointed out above, at each $\mathbf{X}$
we get $\mathbf{P}\in Hom\left( T_{\mathbf{X}}^{\ast }\mathcal{B}_{0},T_{%
\mathbf{x}}^{\ast }\mathcal{B}\right) $ and $\mathcal{S}\in Hom\left( T_{%
\mathbf{X}}^{\ast }\mathcal{B}_{0},T_{\mathbf{w}}^{\ast }\mathbb{V}%
_{w}\right) $.

We now require the \emph{invariance} of $\mathcal{P}_{\mathfrak{b}%
}^{ext}\left( \mathbf{\dot{x},\dot{w}}\right) $ with respect to classical
changes of observers ruled by $SO\left( 3\right) $. For such changes, the
time parametrized family of automorphisms acting on the ambient space $%
\mathcal{E}^{3}$ is the one of isometries so that, as usual, if $\mathbf{%
\dot{x}}^{\mathbf{\ast }}$ is the value of the velocity $\mathbf{\dot{x}}$\ 
\emph{after} the change of observer, we have%
\begin{equation}
\mathbf{\dot{x}}^{\mathbf{\ast }}=\mathbf{c}\left( t\right) +\mathbf{\dot{q}}%
\left( t\right) \wedge \left( \mathbf{x-x}_{0}\right) +\mathbf{\dot{x},}
\end{equation}%
where $\mathbf{c}\left( t\right) $ is the translational velocity, constant
in space, $\mathbf{x}_{0}$ a point chosen arbitrarily and $\mathbf{\dot{q}}%
\wedge \in \mathfrak{so}\left( 3\right) $\ at each $t$. Moreover, still for
such changes of observers, $SO\left( 3\right) $ itself acts also over $%
\mathbb{V}_{w}$\ and we indicate with $\mathbf{\dot{w}}^{\ast }$\ the rate $%
\mathbf{\dot{w}}$ measured after the change of observer, we get%
\begin{equation}
\mathbf{\dot{w}}^{\ast }=\mathbf{\dot{w}+\dot{q}}\left( t\right) \wedge 
\mathbf{w.}
\end{equation}

Then, the \emph{requirement of invariance} is%
\begin{equation}
\mathcal{P}_{\mathfrak{b}}^{ext}\left( \mathbf{\dot{x}}^{\ast }\mathbf{,\dot{%
w}}^{\ast }\right) =\mathcal{P}_{\mathfrak{b}}^{ext}\left( \mathbf{\dot{x},%
\dot{w}}\right)  \label{Ax}
\end{equation}%
for \emph{any} choice of translational $\mathbf{c}$ and rotational $\mathbf{%
\dot{q}}$\ velocities and for any part $\mathfrak{b}$\ (see [M] for a more
general setting involving abstract morphological descriptors).

The arbitrariness of $\mathbf{c}$ and $\mathbf{\dot{q}}$\ and their
independence of space imply from (\ref{Ax}) the integral balances%
\begin{equation}
\int_{\mathfrak{b}}\mathbf{\bar{b}}d\left( vol\right) +\int_{\partial 
\mathfrak{b}}\mathbf{Pn}d\left( area\right) =0,  \label{F}
\end{equation}%
\begin{equation}
\int_{\mathfrak{b}}\left( \left( \mathbf{x-x}_{0}\right) \wedge \mathbf{\bar{%
b}+w}\wedge \mathbf{\beta }\right) d\left( vol\right) +\int_{\partial 
\mathfrak{b}}\left( \left( \mathbf{x-x}_{0}\right) \wedge \mathbf{Pn+w}%
\wedge \mathcal{S}\mathbf{n}\right) d\left( area\right) =0,  \label{Cou}
\end{equation}%
which are the standard integral balance of forces and a non-standard (due to
the presence of the densities of phason interactions) integral balances of
moments. They are the sole global conservation laws associated with the
killing fields of the metric in the ambient space.

The arbitrariness of $\mathfrak{b}$ implies%
\begin{equation}
\mathbf{\bar{b}}+Div\mathbf{P=0}  \label{LF}
\end{equation}%
from (\ref{F}) and%
\begin{equation}
\mathsf{e}\mathbf{PF}^{T}=\mathbf{w}\wedge \left( \mathbf{\beta +}Div%
\mathcal{S}\right) +\left( \nabla \mathbf{w}\right) ^{T}\mathcal{S}
\label{LC}
\end{equation}%
from (\ref{Cou}), with \textsf{e} Ricci's alternating symbol.

The inertial component of $\mathbf{\bar{b}}$, namely $\mathbf{b}^{in}$, and
the explicit expression of $\mathbf{\beta }$\ can be identified by requiring
that their power is the opposite of the rate of the kinetic energy, i.e.%
\begin{equation}
\frac{d}{dt}\left\{ \text{kinetic energy in }\mathfrak{b}\right\} -\int_{%
\mathfrak{b}}\left( \mathbf{b}^{in}\cdot \mathbf{\dot{x}+\beta \cdot \dot{w}}%
\right) d\left( vol\right) =0
\end{equation}%
for any choice of $\mathfrak{b}$\ and of the velocity fields. When
sound-like modes appear in phason activity so that the kinetic energy in $%
\mathfrak{b}$ is given by $\frac{1}{2}\int_{\mathfrak{b}}\left( \rho
_{0}\left\vert \mathbf{\dot{x}}\right\vert ^{2}+\bar{\rho}\left\vert \mathbf{%
\dot{w}}\right\vert ^{2}\right) d\left( vol\right) $, as in IIC, the
arbitrariness of $\mathfrak{b}$\ and of the velocity fields implies%
\begin{equation}
\mathbf{b}^{in}=-\rho _{0}\mathbf{\ddot{x},}\text{ \ \ \ \ \ \ }\mathbf{%
\beta }=-\bar{\rho}\mathbf{\ddot{w}.}
\end{equation}

In this way\ (\ref{LF}) reduces formally to (\ref{Cauchy}). Moreover, from (%
\ref{LC}) we get two information:

\begin{enumerate}
\item at each $\mathbf{X}\in \mathcal{B}_{0}$, the term $\mathsf{e}\mathbf{PF%
}^{T}-\left( \nabla \mathbf{w}\right) ^{T}\mathcal{S}$ is given by the cross
product between $\mathbf{w}$ and an element of $T_{\mathbf{w}}^{\ast }%
\mathbb{V}_{w}\simeq \mathbb{R}^{3}$ that we indicate with $\mathbf{z}$;

\item $\mathbf{z}$ is just equal to $Div\mathcal{S}-\bar{\rho}\mathbf{\ddot{w%
}}$ so that we get (\ref{SI}).
\end{enumerate}

\textbf{Remark 7}. In summary, when we represent interactions due to phason
activity in quasiperiodic crystalline structures, an internal self-force
arises a priori just as a consequence of requirements of $SO\left( 3\right) $
invariance of the power. Constitutive issues render explicit its structure
as a function of state. In this way we find that the conservative part of $%
\mathbf{z}$ disappear for IQ because the relevant elastic energy does not
depend on $\mathbf{w}$ while its dissipative part may play a role as in (\ref%
{Min}) (see [RoLo]).

\section{Dissipative evolution of sharp interfaces}

We consider now the case in which the motion of $\Sigma $ described by the
vector field $\mathbf{\tilde{\varpi}}$ is not virtual, rather it is the 
\emph{real} irreversible motion of $\Sigma $ which can be e.g. identified
with the boundary of an evolving defect. We assume that other irreversible
phenomena like phason or gross friction do not occur. Dissipation is
associated just with the evolution of $\Sigma $ that we presume also to be
coherent in the sense specified in Section 5.

We then introduce a dissipative \emph{surface driving force} $\mathbf{f}%
_{\Sigma }$ along $\Sigma $. Coherence allows us to write the condition of
dissipativity just on the normal component so that we require%
\begin{equation}
\left( \mathbf{f}_{\Sigma }\cdot \mathbf{m}\right) U\leq 0,  \label{Dissip}
\end{equation}%
for any choice of $U$. We then modify (\ref{IntSuf}) by adding a surface
source term of the type%
\begin{equation}
\int_{\mathfrak{b}_{\Sigma }\cap \Sigma }\mathbf{f}_{\Sigma }\cdot \mathfrak{%
w}d\left( area\right) .
\end{equation}

\ \ \ \ \ \ \ \ \ \ \ \ \ \ \ \ \ \ \ \ \ \ 

\textbf{Proposition 1}. \emph{Let} $\Sigma $ \emph{be a structured surface
with surface energy} $\phi $. \emph{Let us assume that during the
dissipative evolution of }$\Sigma $%
\begin{equation*}
\frac{d}{dt}\int_{\mathfrak{b}_{\Sigma }}\mathcal{Q}d\left( vol\right)
+\int_{\partial \mathfrak{b}_{\Sigma }}\mathfrak{F}\cdot \mathbf{n}d\left(
area\right) +
\end{equation*}%
\begin{equation}
+\int_{\partial \left( \mathfrak{b}_{\Sigma }\cap \Sigma \right) }\mathfrak{X%
}\cdot \mathsf{n}d\left( length\right) +\int_{\mathfrak{b}_{\Sigma }\cap
\Sigma }\mathbf{f}_{\Sigma }\cdot \mathfrak{w}d\left( area\right) =0
\end{equation}%
\emph{for any part} $\mathfrak{b}_{\Sigma }$ \emph{of} $\mathcal{B}_{0}$ 
\emph{crossing} $\Sigma $. \emph{If} $\mathcal{L}$ \emph{and} $\phi $\ \emph{%
are invariant in the sense of Theorem 5 and }$\mathbf{\hat{f}}_{s_{1}}^{1}$%
\emph{\ acts alone, the evolution of }$\Sigma $ \emph{along its normal is
ruled by} 
\begin{equation*}
\mathbf{m\cdot }\left[ \mathbb{P}\right] \mathbf{m}+\mathbb{C}_{\tan }\cdot 
\mathsf{L}+Div_{\Sigma }\mathfrak{c}+
\end{equation*}%
\begin{equation}
+\frac{1}{2}\rho _{0}U^{2}\left[ \left\vert \mathbf{Fm}\right\vert ^{2}%
\right] -\bar{\rho}U\left[ \left( \nabla \mathbf{w}\right) ^{T}\mathbf{\dot{w%
}}\right] \cdot \mathbf{m}-\frac{1}{2}\bar{\rho}\left[ \left\vert \mathbf{%
\dot{w}}\right\vert ^{2}\right] =\tilde{f}_{\Sigma }U,  \label{DiE}
\end{equation}%
\emph{where} $\tilde{f}_{\Sigma }$ \emph{is a positive driving coefficient
such that }$\tilde{f}_{\Sigma }=\hat{f}_{\Sigma }\left( \mathbf{m},U\right) $%
.

\ \ \ \ \ \ \ \ \ \ \ \ \ \ 

The proof follows Theorem 5 basically. In addition one may realize that the
condition (\ref{Dissip}) implies that the normal component $f_{\Sigma }=%
\mathbf{f}_{\Sigma }\cdot \mathbf{m}$ be of the form $f_{\Sigma }=-\tilde{f}%
_{\Sigma }U$ with $\tilde{f}_{\Sigma }$ a positive coefficient.

\textbf{Remark 8}. In absence of phason activity, (\ref{DiE}) reduces to the
balance equation describing the dissipative evolution of surfaces in simple
bodies as discussed in [G] in presence of bulk deformation and surface
energy. When bulk deformation is also absent, we may get the anisotropic
motion by curvature.

\ \ \ \ \ \ \ \ \ \ \ \ \ \ \ \ \ 

\textbf{Acknowledgement}. The support of the Italian National Group of
Mathematical Physics (GNFM-INDAM) is acknowledged.

\section{References}

\begin{description}
\item[{[A]}] Antman, S. S. (1998), Physically unacceptable viscous stresses, 
\emph{Z. angew. Math. Phys.}, \textbf{49}, 980-988.

\item[{[BH]}] Baake, M. and H\"{o}ffe, M. (2000), Diffraction of random
tilings: some rigorous results, \emph{J. Stat. Phys.}, \textbf{99}, 219-261.

\item[{[BKMR]}] Bloch, A., Krishnaprasad, P. S., Marsden, J. E., Ratiu, T.
(1996), The Euler-Poincar\'{e} equations and double bracket dissipation, 
\emph{Comm. Math. Phys.}, \textbf{174}, 1-42.

\item[{[C]}] Capriz, G., \emph{Continua with microstructure}.
Springer-Verlag, Berlin, 1989.

\item[{[CM]}] Capriz, G., Mariano, P. M. (2003), Symmetries and Hamiltonian
formalism for complex materials, \emph{J. Elasticity}, \textbf{72}, 57-70.

\item[{[CM98]}] Miehe, C., A formulation of finite elastoplasticity based on
dual co- and contra-variant eigenvector triads normalized with respect to a
plastic metric, \emph{Comp. Meth. Appl. Mech. Eng.}, \textbf{159}, 223-260.

\item[{[D]}] Del Piero, G. (2003), A class of fit regions and a universe of
shapes for continuum mechanics, \emph{J. Elasticity}, \textbf{70}, 175-195.

\item[{[dFM]}] de Fabritiis, C., Mariano, P. M. (2004), Geometry of
interactions in complex bodies, \emph{submitted} (arXiv: math-ph/0406036).

\item[{[DP]}] De, P., Pelcovits, R. A. (1987), Linear elasticity theory of
pentagonal quasicrystals, \emph{Physical Review} B, \textbf{35}, 8609-8620.

\item[{[DYHW]}] Ding, D.-H., Yang, W., Hu, C., Wang, R. (1993), Generalized
theory of quasicrystals, \emph{Physical Review} B, \textbf{48}, 7003-7010.

\item[{[G]}] Gurtin, M. E., \emph{Configurational forces as basic concepts of
continuum physics}, Springer-Verlag, New York, 2000.

\item[{[GC]}] Capriz, G. (1985), Continua with latent microstructure, \emph{%
Arch. Rational Mech. Anal.}, \textbf{90}, 43-56.

\item[{[HWD]}] Hu, C., Wang, R. and Ding, D.-H. (2000), Symmetry groups,
physical property tensors, elasticity and dislocations in quasicrystals, 
\emph{Rep. Prog. Phys.}, \textbf{63}, 1-39.

\item[{[JS]}] Jeong, H.-C., Steinhardt, P. J. (1993), Finite-temperature
elasticity phase transitions in decagonal quasicrystals, \emph{Physical
Review} B, \textbf{48}, 9394-9403.

\item[{[L]}] Lifshitz, R. (2003), Quasicrystals: a matter of definition, 
\emph{Found. Phys.}, \textbf{33}, 1703-1711.

\item[{[LRT]}] Lubensky, T. C., Ramaswamy, S., Toner, J. (1985),
Hydrodynamics of icosahedral quasicrystals, \emph{Physical Review} B, 
\textbf{32}, 7444-7452.

\item[{[M]}] Mariano, P. M. (2001), Multifield theories in mechanics of
solids, \emph{Adv. Appl. Mech.}, \textbf{38}, 1-93.

\item[{[M03]}] Mariano, P. M. (2003), Cancellation of vorticity in
steady-state non-isentropic flows of complex fluids, \emph{J. Phys. A Math.
Gen.}, \textbf{36}, 9961-9972.

\item[{[M04]}] Mariano, P. M., \emph{Elements of multifield theories for
complex bodies}, Birkhauser (Springer) Verlag, Boston, in preparation.

\item[{[M1]}] Mariano, P. M. (2004), Consequences of `changes' of material
metric in simple bodies, \emph{Meccanica}, \textbf{39}, 77-79.

\item[{[MA]}] Mariano, P. M., Augusti, G. (1998), Ordering and transport in
generalized continua with arbitrary order parameters, \emph{J. Phys. IV}, 
\textbf{8}, 223-230.

\item[{[ML]}] Mandal, R. K., Lele, S. (2000), Interfaces in quasicrystals:
problems and prospects, \emph{Mat. Sci. Eng. A - Struct.}, \textbf{294},
813-817.

\item[{[MR]}] Marsden, J. E., Ratiu, T., \emph{Introduction to mechanics and
symmetry}, Springer-Verlag, New York, 1999.

\item[{[MSA]}] Mariano, P. M., Stazi, F. L., Augusti, G. (2004), Phason
effects around a crack in Al-Pb-Mn quasicrystals: stochastic aspects of the
phonon-phason coupling, \emph{Computers \& Structures}, \textbf{82}, 971-983.

\item[{[RL]}] Rochal, S. B. and Lorman, V. L. (2000), Anisotropy of
acoustic-phonon properties of an icosahedral quasicrystal at high
temperature due to phonon-phason coupling, \emph{Physical Review }B, \textbf{%
62}, 874-879.

\item[{[RoLo]}] Rochal, S. B., Lorman, V. L. (2002), Minimal model of the
phonon-phason dynamics in icosahedral quasicrystals and its application to
the problem of internal friction in the \emph{i}-AlPbMn alloy, \emph{%
Physical Review} B, \textbf{66}, 144204 (1-9).

\item[{[RT]}] Richer, M., Trebin, H.-R. (2002), Non-linear generalized
elasticity of icosahedral quasicrystals, \emph{J. Phys. A Math. Gen.}, 
\textbf{35}, 6953-6962.

\item[{[S]}] Shechtman, D., Blech, I., Gratias, D., Cahn, J. W. (1984),
Metallic phase with long-range orientational order and no translational
symmetry, \emph{Phys. Rev. Letters}, \textbf{53}, 1951-1954.

\item[{[Si]}] \v{S}ilhav\'{y}, M., \emph{The mechanics and thermodynamics of
continuous media}, Springer, Berlin, 1997.
\end{description}

\end{document}